\documentclass[a4paper,11pt]{article}
\usepackage{float,jheppub,slashed} 
\usepackage{lineno}
\usepackage{subfigure}
\usepackage{makecell}

\arxivnumber{2609.05900} 

\title{\boldmath Prospective study of light dark matter search in positron beam mode with DarkSHINE experiment initiative}

\author[1,2]{Xu-Liang Zhu,}
\author[1,2]{Haidar Mas'ud Alfanda,}
\author[1,2]{Shu Li,}
\author[2,1]{Kun Liu,}
\author[1,2]{Tong Sun,}
\author[2,1]{Haijun Yang,}
\author[1,2]{Yi-Fan Zhu}
\affiliation[1]{\it State Key Laboratory of Dark Matter Physics, Key Laboratory for Particle Astrophysics and Cosmology (KLPPAC-MoE), Shanghai Key Laboratory for Particle Physics and Cosmology (SKLPPC), Tsung-Dao Lee Institute, Shanghai Jiao Tong University, Shanghai 201210, China}
\affiliation[2]{\it State Key Laboratory of Dark Matter Physics, Key Laboratory for Particle Astrophysics and Cosmology (KLPPAC-MoE), Shanghai Key Laboratory for Particle Physics and Cosmology (SKLPPC), Institute of Nuclear and Particle Physics, School of Physics and Astronomy, Shanghai Jiao Tong University, Shanghai 200240, China}

\abstract{We perform the prospective study on searching for invisibly decayed dark photons ($A^{\prime}$) in positron-on-fixed-target (POT) scheme with the DarkSHINE experiment simulation setup. In the positron-on-fixed-target scheme, two additional production channels of dark photons via annihilation with extranuclear electrons are present: non-resonant t-channel ($e^+e^-N \to \gamma A^{\prime}N$) and resonant s-channel ($e^+e^-N \to A^{\prime}N$), alongside bremsstrahlung emission similar to electron-on-fixed-target scheme. Based on a GEANT4 full detector simulation of an 8 GeV positron beam with $3 \times 10^{14}$ POT events, we developed an analysis strategy using missing energy and momentum signatures across three orthogonal signal regions optimized for these channels. Our results indicate that the additional annihilation channels of dark photon production enhance the dark photon invisible decay search sensitivity, improving the exclusion limits by approximately two orders of magnitude compared to previous constraints in the electron-on-fixed target (EOT) scheme, particularly near the center-of-mass energy threshold for annihilation production of dark photons.}

\makeatletter
\def\@fpheader{\relax}
\makeatother
\begin{document}
\maketitle
\flushbottom

\section{Introduction}
\label{sec:intro}

The Standard Model (SM) explains the fundamental particles and forces that make up the universe, but it only explains about $5\%$ of total matter. The other $95\%$ come from dark energy and dark matter (DM), which remain unknown in the beyond SM (BSM) scenarios~\cite{Bertone:2016nfn,Battat:2024mcc}. Although the existence of DM has been proved by numerous astronomical and cosmological observations, and it interacts gravitationally with SM particles, its microscopic properties at elementary particle level remain a profound mystery~\cite{Battat:2024mcc,Hooper:2007kb,Liddle:1998ew,Clowe:2006eq}. 
For decades in both theoretical and experimental particle physics, the searches for DM have been thoroughly carried out in almost all frontiers of particle physics including cosmic frontier, intensity frontier and energy frontier. Being one of the primarily focused DM benchmark model, Weakly Interacting Massive Particles (WIMPs) are considered one of the most well-motivated candidates for DM. However, despite an extensive search for direct detection experiments, indirect search experiments, and collider experiments, no definitive signals have been observed~\cite{Arcadi:2017kky}. Various experimental upper limits have ruled out a lot of previously favored parameter space, particularly in the GeV-TeV mass range, leaving the sub-GeV region as a critical frontier for exploration~\cite{Billard:2021uyg}. 

In many BSM models, the Light Dark Matter (LDM) hypothesis assumes that DM is made up of sub-GeV particles, interacting with SM through new interactions. Among those hypothetical BSM mediators, dark photon ($A^{\prime}$) is one of the popular vector portal mediators in many simplified BSM scenarios, which can kinetically mix with the ordinary photon, parametrized with a kinetic mixing parameter $\epsilon$, and connecting DM and the SM particles through the new BSM interactions that it carries~\cite{Holdom:1985ag,Fuyuto:2019vfe,Cheng:2021qbl,Choi:2020dec}. 

\begin{figure}[htbp]
\centering
\subfigure[$e^+N \to e^+A^{\prime}N$]{
    \includegraphics[width=.3\textwidth]{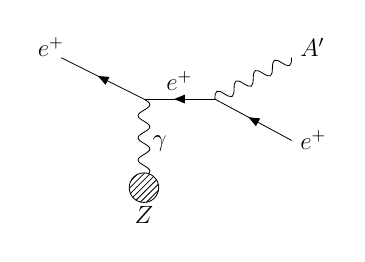}}
\subfigure[$e^+e^-N \to \gamma A^{\prime}N$]{
    \includegraphics[width=.3\textwidth]{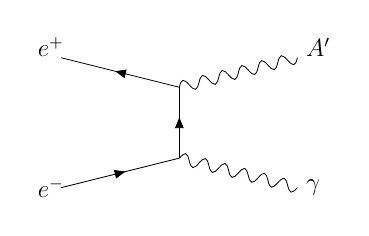}}
\subfigure[$e^+e^-N \to A^{\prime}N$]{
    \includegraphics[width=.3\textwidth]{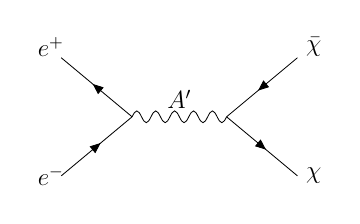}}
\caption{Three different $A^{\prime}$ production mechanisms by positrons on a fixed target $Z$: (a) \textit{bremsstrahlung} $A^{\prime}$ emission; (b) t-channel $A^{\prime}$ production. (c) s-channel $A^{\prime}$ production. Note that in process (b) and (c), $N$ in the expression is to indicate the $e^-$ is from the target material, and target nucleus is not included in the vertex calculation of this two process. }
\label{fig:darkphoton}
\end{figure}

Currently, many international LDM search experiments and initiatives such as NA64~\cite{Banerjee:2019pds,NA64:2023wbi}, LDMX~\cite{LDMX:2019gvz}, and DarkSHINE~\cite{DarkSHINE:2022mak} have extensively explored dark photons using electron-on-fixed-target (EOT) schemes. Among these, the DarkSHINE experiment initiative, which aims to utilize the high-repetition rate single-electron beam generated by the Shanghai High Repetition-Rate XFEL and Extreme Light Facility (SHINE) to detect LDM, shows that its promising protentatial being sensitive to LDM, especially for dark matter below 100 MeV via dark photon invisible decay searches~\cite{DarkSHINE:2022mak}. However, the prospective search sensitivity in positron-on-fixed-target (POT) scheme has not yet been studied previously in the DarkSHINE experiment setup. The dark photon can be generated in the POT scheme via three processes in \Cref{fig:darkphoton}, with the final state $A^{\prime}$ decaying to a pair of DM particles ($\chi\bar\chi$)~\cite{Battaglieri:2021rwp,Marsicano:2018glj}. Compared to EOT scheme, the $A^{\prime}$ can be generated also through positron annihilation with extranuclear electrons, besides the dark bremsstrahlung $A^{\prime}$ production similar to the EOT scheme. The effect of $A^{\prime}$ t-channel production has been discussed in fixed target experiments with positron beams in Ref.~\cite{Raggi:2014zpa,Wojtsekhowski:2017ijn,Alexander:2017rfd}. In addition, the limits of the $A^{\prime}$ parameter space in the 200-300 MeV region were extended by almost an order of magnitude by including $A^{\prime}$ production via secondary positron annihilation with extranuclear electrons for electron beam in NA64~\cite{Andreev:2021fzd}. Moreover, the improvement of the existing exclusion limit by s-channel via positron beam has been discussed for NA64 in Ref.~\cite{NA64:2023ehh}. 

In this work, we focus on studying the search sensitivity in the POT mode for dark photon invisible decay signal with the DarkSHINE experiment simulation setup the same as the simulation analysis in electron beam mode as shown in our previous work~\cite{DarkSHINE:2022mak}. Despite the positron beam is introduced only hypothetically in this study, it aims to conclude on quantified and concrete sensitivity that can potentially motivate the dark photon search experimental program with more diversified beamline approaches with the SHINE facility. In the POT scheme, the dark photons are generated by dark bremsstrahlung and positron annihilation processes in \Cref{fig:darkphoton}. The missing-energy strategy, noted for its effectiveness in the search for dark photons, as referenced in~\cite{Izaguirre:2014bca}, is employed in these three processes. In this simulation analysis, 2.5 billion inclusive POT events and dedicated samples for each of major backgrounds are simulated with GEANT4. Finally, the overall limits are determined by considering the three Feynman diagrams contributing to three signal regions. By considering the contribution of positron annihilation through t-channel ($e^+e^-N \to \gamma A^{\prime}N$) and s-channel ($e^+e^-N \to A^{\prime}N$) production, the exclusion limits are pushed downwards in certain kinematic regions, improving the experimental search sensitivity with respect to the EOT scheme experiment given comparable luminosities.

\section{Theoretical assumptions} \label{ch:theory}

One of the most well-motivated extensions of the Standard Model (SM) to incorporate new interactions with DM is the introduction of an additional light $U(1)_D$ gauge boson, $A^{\prime}$, commonly referred to as the dark photon. The dark photon can mix with the SM photon through kinetic mixing, thereby acting as a portal to the dark sector. In addition to the minimal $U(1)_D$ extension, we consider an interaction term between the dark fermion $\chi$ and $A^{\prime}$. The resulting extended Lagrangian is given by:

\begin{equation}
\mathcal{L} = \mathcal{L}_{\text{SM}} - \frac{1}{4} F_{\mu\nu}^{\prime} F^{\prime\mu\nu} + \frac{1}{2} m_{A^{\prime}}^{2}  A_{\mu}^{\prime}A^{\prime\mu} -  \frac{\varepsilon}{2}  F_{\mu\nu} F^{\prime\mu\nu} - \sqrt{4 \pi \alpha_D} \bar{\chi}\gamma^{\mu
}\chi A^{\prime}_{\mu} ,
\end{equation}

where the mass term of the dark fermion $\chi$ is omitted. Here, $F^{\prime}_{\mu\nu} = \partial_\mu A^{\prime}_\nu - \partial_\nu A^{\prime}_\mu$ is the field strength tensor of the dark photon, $m_{A^{\prime}}$ is the dark photon mass, $\varepsilon$ is the kinetic mixing strength, $F_{\mu\nu}$ is the electromagnetic tensor, and $\alpha_D$ is the coupling constant associated with the $U(1)_D$ gauge group. If $m_\chi < \frac{1}{2}m_{A^{\prime}}$, the decay width of $A^{\prime} \to \chi\bar{\chi}$ is proportional to $\alpha_D$. Electroweak symmetry breaking (EWSB) introduces an interaction term between SM fermions and the dark photon: 

\begin{equation}
\mathcal{L}_{int, A^{\prime}} = - \epsilon e \bar{\psi}\gamma^{\mu}\psi A^{\prime}_{\mu} ,
\end{equation}

The decay width of $A^{\prime} \to l^+l^-$ or $A^{\prime} \to \text{hadrons}$ is proportional to $\alpha \epsilon^2$. The Dark SHINE experiment is designed to be particularly sensitive to scenarios where $\alpha_D \gg \alpha \epsilon^2$, which is an interesting possibility that it suppresses the decay channels of the dark photon into visible particles, making the dominant decay channel the invisible decay $A^{\prime} \to \chi\bar{\chi}$.  

At tree level, it is typically assumed that $\sqrt{4 \pi \alpha_D} \sim 1$. The parameter $\varepsilon$ is expected to lie in the range $\sim 10^{-4} - 10^{-2}$ (or $\sim 10^{-6} - 10^{-3}$) if the mixing is generated by one-loop (or two-loop) interactions~\cite{Holdom:1985ag,Essig:2010ye,delAguila:1988jz,Arkani-Hamed:2008kxc}. In this scenario, the production signature of $A^{\prime}$ consists of a large missing energy.  

\begin{figure}[htbp]
    \centering
    \includegraphics[width=0.5\linewidth]{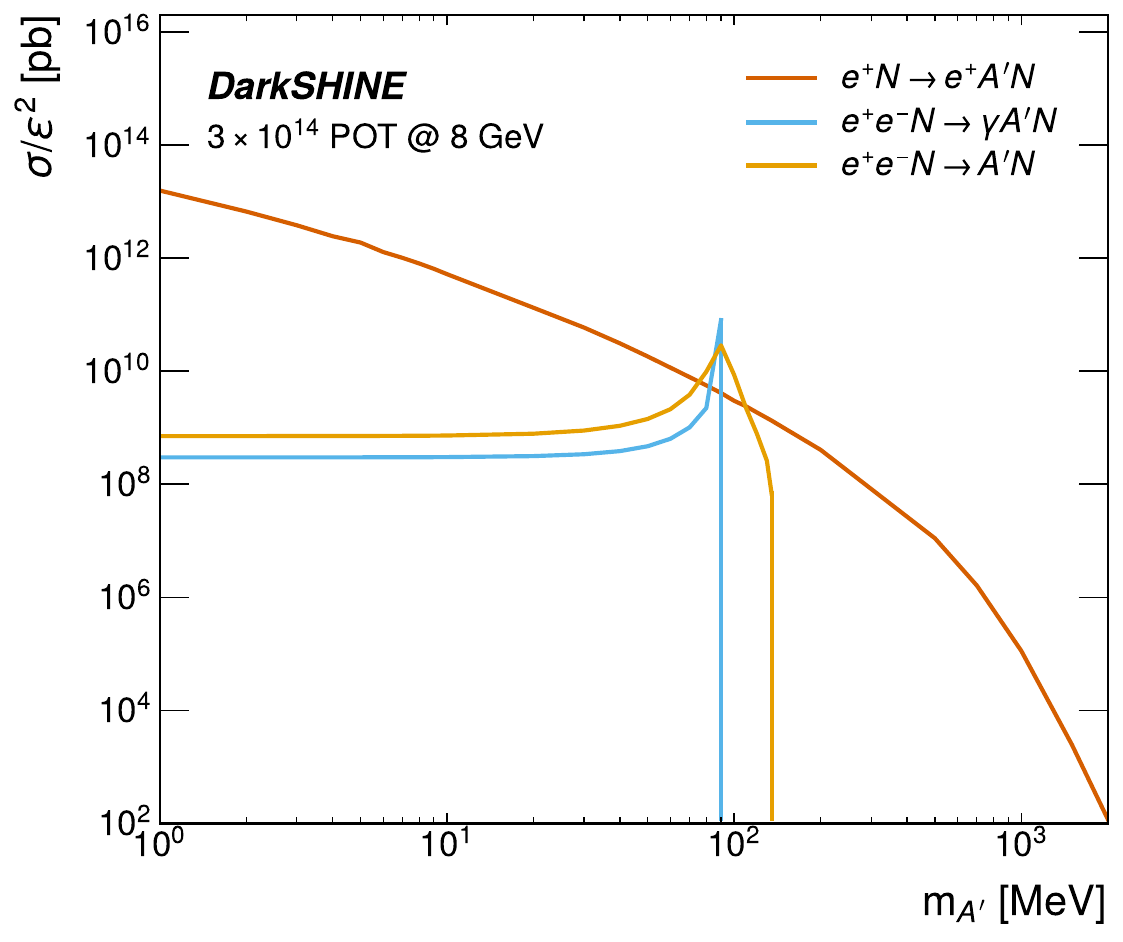}
    \caption{Inclusive cross sections for dark photons via the three channels: \textit{bremsstrahlung} ($e^+N \to e^+A^{\prime}N$), t-channel production ($e^+e^-N \to \gamma A^{\prime}N$), and s-channel production ($e^+e^-N \to A^{\prime}N$). The cross sections are produced by CalcHEP\cite{Belyaev:2012qa} with 8 GeV $e^+$ beam energy. The sharp stopping of t-channel here is caused by we only consider on-shell final state of $A^{\prime}$.}
    \label{fig:crossSection}
\end{figure}

In positron-on-target experiments, dark photons ($A^{\prime}$) are produced through three distinct channels: \textit{Bremsstrahlung} emission ($e^+N \to e^+NA^{\prime}$),  t-channel production ($e^+Ne^- \to \gamma A^{\prime}N$), and s-channel production ($e^+Ne^- \to A^{\prime}N$). The cross sections for these processes are derived from the kinetic mixing model and the dark photon couples to SM particles via the mixing parameter $\epsilon$. These cross sections depend critically on the dark photon mass $m_{A^{\prime}}$ and the beam energy. \Cref{fig:crossSection} illustrates the cross sections (normalized by $\varepsilon^2$) as functions of $m_{A^{\prime}}$ for an 8 GeV $e^+$ beam energy.  

\subsection{Bremsstrahlung Production}
The dominant process at low $A^{\prime}$ masses ($m_{A^{\prime}} \ll \sqrt{s}$) is the \textit{bremsstrahlung} emission of $A^{\prime}$ from positrons scattering off nuclei in the target. The differential cross section which is proportional to the square of the mixing parameter $\epsilon$ is given by~\cite{Andreas:2012mt}:
\begin{equation}
\frac{d\sigma_{brem}}{x} = 4\pi\varepsilon^2 \alpha^3 \xi \sqrt{1-\frac{m_{A^{\prime}}^2}{E_{e^+}}} \frac{1 - x + \frac{x^2}{3}}{m_{A^{\prime}}^2\frac{1-x}{x} + m_e^2x},
\label{eq:brems}
\end{equation}
where $x_e \equiv E_{A^{\prime}}/E_{e^+}$, $\alpha$ is the fine-structure constant, and $\xi$ is the effective flue of photons which is associated with the electric form factor~\cite{Bjorken:2009mm}.

\subsection{t-channel Production}
For higher mass regions, channel $e^+e^-N \to \gamma A^{\prime}N$ contributes significantly. The total cross section for this process is given by~\cite{Marsicano:2018krp}:
\begin{equation}
\sigma_{t} = \frac{8\alpha^3 \varepsilon^2}{s^2} \left[\left(\frac{s-m_{A^{\prime}}^2}{2s} + \frac{m_{A^{\prime}}^2}{s-m_{A^{\prime}}^2}\right)log\frac{s}{m_e^2} -  \frac{s-m_{A^{\prime}}^2}{2s}\right],
\label{eq:nonres}
\end{equation}
where $s$ is the $e^+ e^-$ system invariant mass squared, $m_e$ is positron mass. This process becomes particularly relevant when $m_{A^{\prime}}$ approaches the kinematic threshold of the experiment.

\subsection{s-channel Production}
When the collision energy matches the dark photon mass ($\sqrt{s} \approx m_{A^{\prime}}$), resonant annihilation $e^+e^- \to A^{\prime} \to \chi\bar{\chi}$ dominates. The cross section for a vector $A^{\prime}$ decaying into LDM is given by~\cite{Andreev:2021fzd}:  
\begin{equation}  
\sigma_{\text{s}} = \frac{4\pi\alpha \alpha_D \varepsilon^2}{\sqrt{s}} \frac{q\kappa}{(s - m_{A^{\prime}}^2)^2 + m_{A^{\prime}}^2\Gamma_{A^{\prime}}^2\eta},  
\label{eq:res}  
\end{equation}  
where $q$ is the momentum of the LDM daughter particles in the center-of-mass (CM) frame, $\Gamma_{A^{\prime}}$ is the width of $A^{\prime}$, given by $\Gamma_{A^{\prime}} = \alpha_D \frac{m_{A^{\prime}}}{3}(1 + 2r^2)\sqrt{1 - 4r^2}$ for fermionic LDM, and $r = m_\chi/m_{A^{\prime}}$. The kinematic factor $\kappa$ is defined as $s - \frac{4}{3}q^2$ ($\frac{2}{3}q^2$) for the fermionic (scalar) case, while $\eta = (s/m_{A^{\prime}}^2)^2$ is a correction term introduced for the fermionic LDM case ($\alpha_D = 0.5$) to account for the energy dependence of $\Gamma_{A^{\prime}}$ when it is non-negligible compared to $m_{A^{\prime}}$.  

\section{Detector and Simulation Framework}

The SHINE facility is currently under construction and close to completion~\cite{Zhao:2018lcl, Wan:2022het}. The previous work~\cite{DarkSHINE:2022mak}, which we studied with a conceptual design of DarkSHINE experiment~\cite{DarkSHINE:CDR}, the high repetition rate single electron beam that SHINE could feasibly deliver offers good opportunity to achieve competitive dark photon search sensitivity. 
The DarkSHINE detector is designed to search for dark photons using an 8 GeV electron beam (or positron beam in this study) with a repetition rate of up to 10 MHz. For a one-year run, this corresponds to $3 \times 10^{14}$ POT events, which is used as the benchmark statistics in comparison with previous work in the EOT scheme so as to demonstrate the search sensitivity difference in between the two different beam conditions.

The study focuses on the invisible decay channel, and thus the detector is designed to be sensitive and precise in measuring missing momentum and missing energy. To measure the momentum/energy of the incident electron or positron beam, and the particles scattered off the tungsten target, the overall detector system consisting of a tracking system incorporating a magnetic field and silicon micro-strip detector~\cite{DarkSHINE:TrackerLGAD} along with calorimetry system with crystal electromagnetic calorimeter~\cite{DarkSHINE:ECALDesign} and hadronic calorimeter~\cite{DarkSHINE:HCALDesign} are utilized. Compared to the bremsstrahlung process, dark bremsstrahlung, which generates massive dark photons, results in a significantly lower energy and lower momentum of the recoil particle.

\subsection{Detector Configuration}
\begin{figure}[htbp]
  \centering
  \includegraphics[width=0.8\textwidth]{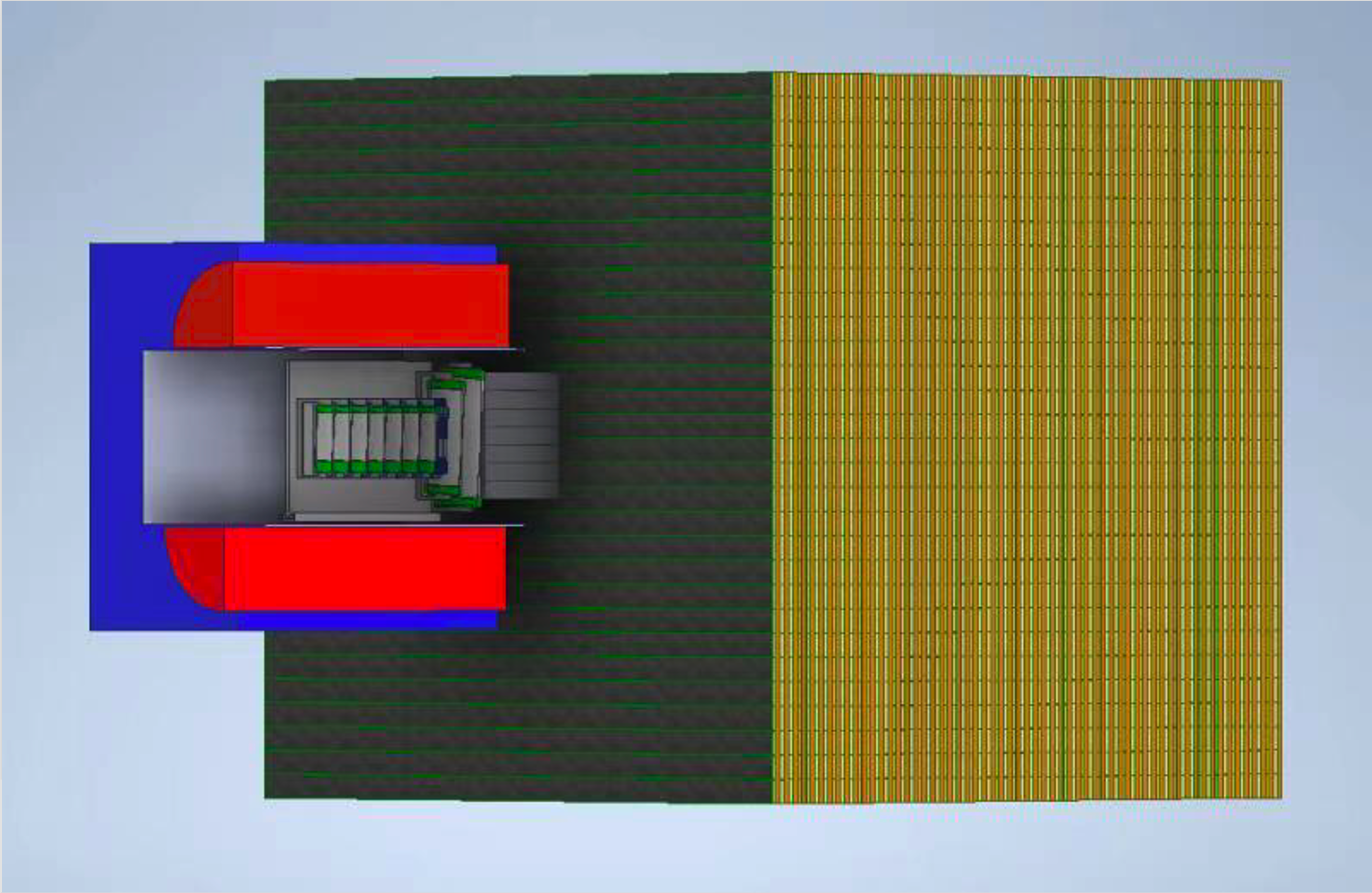}
  \caption{Schematic of DarkSHINE detector layout. Components from upstream to downstream: (1) Tagging tracker, (2) Tungsten target, (3) Recoil tracker, (4) ECAL/HCAL calorimeters.~\cite{DarkSHINE:2022mak}}
  \label{fig:detector_layout}
\end{figure}

The DarkSHINE detector system~\cite{DarkSHINE:TrackerLGAD, DarkSHINE:ECALDesign, DarkSHINE:HCALDesign}, comprises four core components arranged along the beam axis (shown schematically in \Cref{fig:detector_layout}). The \textbf{Tagging Tracker} is a 7 layers of silicon microstrip detector to measure incident positron momenta with precision $\sigma/p \leq$ 3\% at 8 GeV. Immediately lies a 350 $\mu$m tungsten target ($0.1 X_0$ radiation length). The \textbf{Recoil Tracker} system combines 6 layers of silicon detectors within a 0.75 T dipole field, with momentum resolution of $\sigma/p \simeq 6\% $. Energy measurement is performed by two calorimeters: a \textbf{electromagnetic calorimeter (ECAL)} with LYSO crystals ($2.5\times2.5\times4\text{ cm}^2$ cells, $\sigma/E = 0.19/\sqrt{E}\oplus 0.008 \oplus 0.2/E [MeV]$) for precise electromagnetic shower detection, and a \textbf{hadronic calorimeter (HCAL)} using iron/scintillator sampling to veto MIP particles, in particular muons and neutral hadrons.

\subsection{Simulation Framework}
Signal and background processes are simulated using GEANT4 \cite{GEANT4:2002zbu} with FTFP-BERT physics list, and custom physics lists for dark photon. Dark photon production mechanisms include:
\begin{itemize}
    \item Bremsstrahlung: $e^+ N \to e^+ A^{\prime} N$
    \item $e^+e^-$ annihilation: $e^+e^-N\to A^{\prime}N$ and $e^+e^-N\to\gamma A^{\prime}N$
\end{itemize}

Main background sources encompass photonuclear reactions (PN), gamma to muon pairs (GMM), and electro-nuclear interactions (EN).

\section{Signal and background overview}

This section discusses the signal characteristics and background sources for the positron-beam setup in DarkSHINE, and the difference compared to the electron-beam configuration. Compared to the electron-beam setup, the positron-beam configuration introduces two additional production mechanisms for dark photons ($A^{\prime}$) through processes: $e^+e^-N \to \gamma A^{\prime}N$ and $e^+e^-N \to A^{\prime}N$. The signal signature is also missing energy, where the $A^{\prime}$ decays invisibly to light dark matter ($\chi\bar{\chi}$), and the corresponding experimental signature will be discussed in \Cref{sec:sigProduction}. There are several major rare process background sources listed in \Cref{tab:bkg_summary}, including photon–nuclear interactions, electron-positron annihilation, electron–nuclear interactions, and photon decays to muon pairs. Unlike electron beams, positrons annihilate with atomic electrons ($e^+e^- \to \gamma\gamma$ and $e^+e^- \to \gamma\gamma\gamma$), introducing unique backgrounds. This contribution to background also has been take into account for rare processes and will be discussed in \Cref{sec:bkgProduction}

\subsection{Signal sample production}
\label{sec:sigProduction}

The signal distribution was simulated separately for each of the three channels (detailed in Chapter~\ref{ch:theory}) using CalcHEP. Then, the signal samples were produced through a full GEANT4-based simulation with the DarkSHINE detector setup. As shown in \Cref{tab:sig_sample}, there are 30, 18, and 23 signal samples produced for three production mechanisms, each sample with $1\times10^5$ events with a certain dark photon mass.
\begin{table}[htbp]
    \centering
    \caption{Summary of signal samples production}
    \begin{tabular}{c|p{9cm}}
    \hline
    production mechanism of signal & dark photon mass (MeV) \\
    \hline
    \textit{bremsstrahlung} emission & 1, 2, 3, 4, 5, 6, 7, 8, 9, 10, 20, 30, 40, 50, 60, 70, 80, 90, 100, 110, 120, 130, 135, 140, 200, 500, 700, 1000, 1500, 2000\\
    non-resonant annihilation & 1, 2, 3, 4, 5, 6, 7, 8, 9, 10, 20, 30, 40, 50, 60, 70, 80, 90 \\
    resonant annihilation &1, 2, 3, 4, 5, 6, 7, 8, 9, 10, 20, 30, 40, 50, 60, 70, 80, 90, 100, 110, 120, 130, 135 \\
    \hline
    \end{tabular}
    \label{tab:sig_sample}
\end{table}

The incident positron beam scatters off the target, where a dark photon is produced by bremsstrahlung, or through $e^+e^-$ annihilation in the t-channel and s-channel. Based on the detector setup as shown in the \Cref{fig:detector_layout}, the experiment signature will be different for three signal-$A^{\prime}$ production mechanism as discussed in Chapter~\ref{sec:signal_region}. The signature of $e^+N \to e^+A^{\prime}N$ process is similar to the analysis of the electron beam on target as described in detail in Ref.~\cite{DarkSHINE:2022mak} that most of the incident momentum is transferred to dark photon and less than $1/4$ of the incident momentum is carried by recoil electron. Therefore, the difference between the momentum of the tagged track and the recoil track plays an important role in signal region definition. In the case of $e^+e^-N \to \gamma A^{\prime}N$, the $A^{\prime}$ escapes detection, resulting a large difference between the momentum of the tagged track and the recoil track, while the $\gamma$ through the recoil tracker without a track and then can be detected in the downstream ECAL. For $e^+e^-N \to A^{\prime}N$ process, only a $A^{\prime}$ in final state, its energy is not deposited in ECAL, and the corresponding experimental signature is the missing energy. Therefore, the maximum cell energy in ECAL and HCAL can be an effective cut for resonant-$A^{\prime}$ selection.

\subsection{Background sample production}
\label{sec:bkgProduction}

The Standard Model physics simulation of positron scattering on the target is performed using the GEANT4 $\textrm{FTFP-BERT}$ physics list, which is also used for the simulation of particle interactions with the detector. The background is primarily contributed by rare processes, with the three most significant being electron-nuclear, positron-nuclear, and gamma-to-muon-pair processes. To estimate the background with a larger effective positron-on-target (POT) number, Monte Carlo simulations are conducted by biasing the cross-sections of these three physics processes in two distinct detector regions: the Tracker region (including the target) and the ECAL region. The branching ratios, sample sizes, and effective POT numbers of these biased samples are summarized in \Cref{tab:bkg_summary}.

\begin{table}[htbp]
\centering
\caption{Summary of major rare process background}
\label{tab:bkg_summary}
\begin{tabular}{lrrrr}
\hline
Process & Branching ratio & Generated events & Generation Eff. (\%) & POT \\
\hline
PN ECAL & $1.85\times10^{-4}$ & $1\times10^{8}$ & $27.38\%$ & $1.48\times10^{11}$ \\
Annihil Target & $2.93\times10^{-5}$ & $1\times10^{8}$ & $99.00\%$ & $3.38\times10^{12}$ \\
EN ECAL & $4.27\times10^{-6}$ & $1\times10^{8}$ & $1.38\%$ & $3.23\times10^{11}$ \\
GMM ECAL & $1.35\times10^{-6}$ & $1\times10^{8}$ & $27.27\%$ & $2.02\times10^{13}$ \\
PN Target & $1.32\times10^{-6}$ & $1\times10^{8}$ & $6.59\%$ & $4.99\times10^{12}$ \\
EN Target & $5.28\times10^{-7}$ & $1\times10^{8}$ & $1.47\%$ & $2.78\times10^{12}$ \\
GMM Target & $1.39\times10^{-8}$ & $1\times10^{7}$ & $6.57\%$ & $4.73\times10^{13}$ \\
\hline
\end{tabular}
\end{table}  

Most Standard Model events exhibit small missing momentum and small missing ECAL energy, thus they do not contribute significantly to the background. To skip these events irrelevant to the analysis but will cost simulation time and storage, a loose pre-selection is applied globally to the sample production:

\begin{align}
|P_\mathrm{tagging}-P_\mathrm{recoil}| &> 1\ \mathrm{GeV} \\
E_\mathrm{ECAL}^\mathrm{tot} &< 7\ \mathrm{GeV}
\end{align}

\begin{itemize}
\item GMM Target (ECAL): $\gamma\to\mu\bar{\mu}$ with vertex in Target (ECAL) region. Here the $\gamma$ is from hard-bremsstrahlung or annihilation process with $E_\gamma > 2 \text{ GeV}$
\item PN Target (ECAL): $\gamma p\to\text{hadrons}$ in Target (ECAL) region, including quasi-elastic and deep inelastic scattering. $\gamma$ is also from hard-bremsstrahlung or annihilation process with $E_\gamma > 2 \text{ GeV}$
\item EN Target (ECAL): $e^{+}p\to\text{hadrons}$ in Target (ECAL) region. 
\item Annihil Target: $e^+e^-\to \gamma$ in target region, not including the listed rare process above.
\end{itemize}

An inclusive sample (Standard Model physics without event biasing and without BSM physics) is also produced to validate the biased samples. However, due to its limited sample size, it does not cover the full $3\times10^{14}$ POT. Consequently, it is more effective to estimate the background using the biased samples, as they provide a larger effective POT number and thus lower statistical uncertainty.

\section{Signal region definition and signal efficiency}
\label{sec:signal_region}
To maximize sensitivity to dark photon production via three distinct channels ($e^{+}N \to e^{+}A^{\prime}N$, $e^{+}e^{-}N \to \gamma A^{\prime}N$, $e^{+}e^{-}N \to A^{\prime}N$), three orthogonal signal regions are defined in \Cref{tab:signal_regions}, each optimized for specific kinematic features and background suppression strategies.

\begin{table}[htbp]
\centering
\caption{Selection criteria for three orthogonal signal regions. The following symbols are used to define the signal regions: 
\textbf{$N_{trk}^{tag}$} and \textbf{$N_{trk}^{rec}$} represent the number of reconstructed tracks in the tagging and recoil trackers, respectively. 
\textbf{$P_{tag}$} and \textbf{$P_{rec}$} denote the track momentum measured in the tagging and recoil regions. 
\textbf{$E_{ECAL}^{total}$} is the total energy deposited in the Electromagnetic Calorimeter (ECAL), while \textbf{$E_{ECAL}^{MaxCell}$} refers to the maximum energy recorded in a single ECAL cell. 
Similarly, \textbf{$E_{HCAL}^{total}$} and \textbf{$E_{HCAL}^{MaxCell}$} represent the total energy and the maximum single-cell energy deposited in the Hadronic Calorimeter (HCAL), respectively.}
\label{tab:signal_regions}
\begin{tabular}{lccc}
\hline
\textbf{Selection Criteria} & $e^++E_{\mathrm{miss}}$ & $\gamma+E_{\mathrm{miss}}$ & $E_{\mathrm{miss}}$ \\
\hline
$N_{trk}^{tag}$ (Tagging tracks)       & =1            & =1               & =1               \\
$N_{trk}^{rec}$ (Recoil tracks)        & \textbf{=1}            & \textbf{=0}               & \textbf{=0}               \\
$P_{tag} - P_{rec}$ (Missing momentum) & $> 4$ GeV    & --              & --              \\
$E_{ECAL}^{total}$ (Total ECAL energy) & $< 2.5$ GeV  & $< 2$ GeV       & $< 6.6$ GeV     \\
$E_{ECAL}^{MaxCell}$ (Max ECAL cell)   & --           &\textbf{ $\geq 1$ MeV}    & \textbf{$< 1$ MeV}       \\
$E_{HCAL}^{total}$ (Total HCAL energy) & $< 0.1$ GeV  & $< 0.1$ GeV     & $< 0.1$ GeV     \\
$E_{HCAL}^{MaxCell}$ (Max HCAL cell)   & $< 1$ MeV    & $< 1$ MeV       & $< 1$ MeV       \\
\hline
\end{tabular}
\end{table}

\begin{figure}[htbp]
\centering
\includegraphics[width=.45\textwidth]{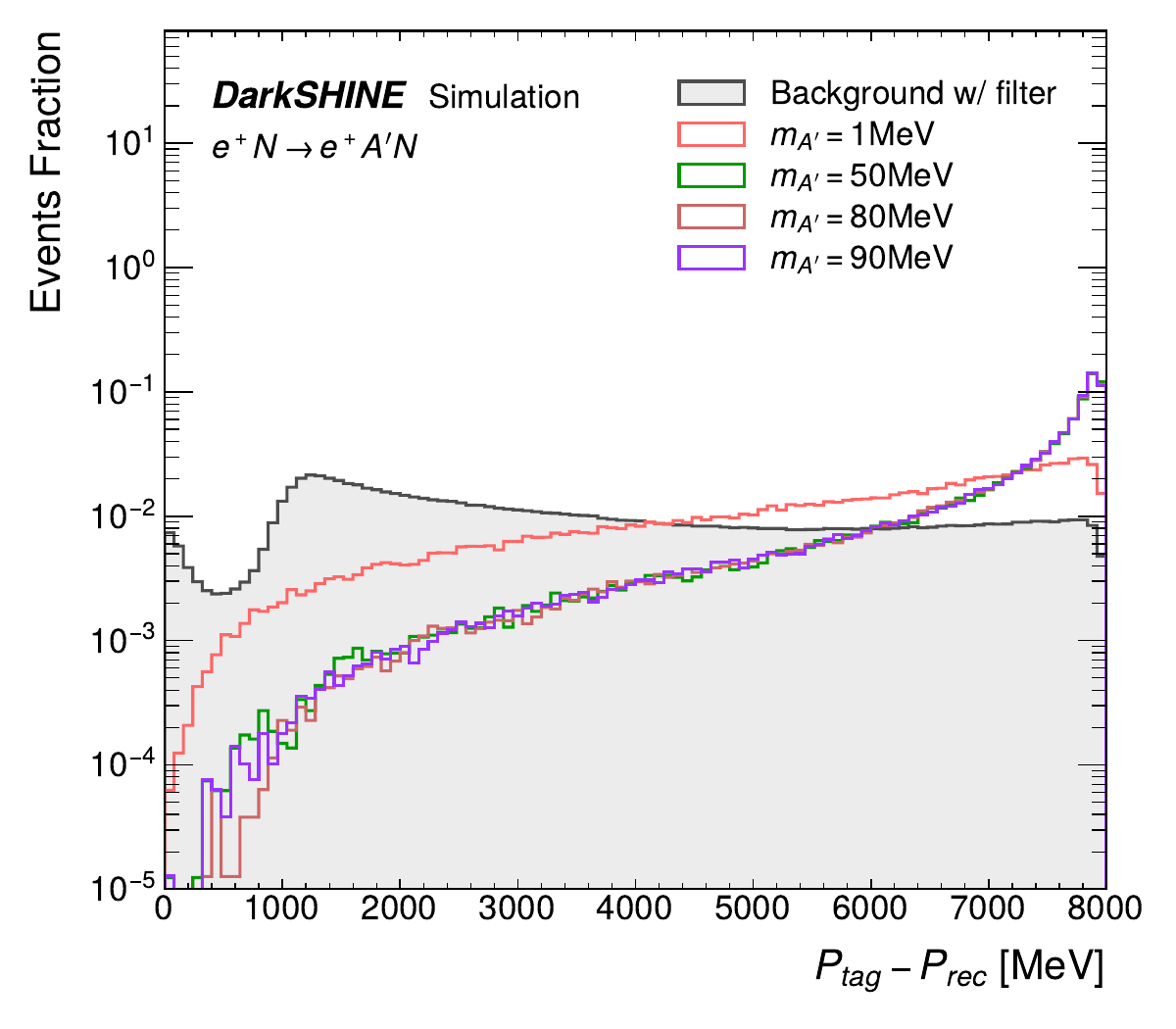}
\caption{The missing momentum $P_{tag}-P_{rec}$ of the (positron) track scatter-off target. The black curve is background distribution with pre-selection $|P_{tag}-P_{rec}|>1 \text{ GeV}$ and $E_{ECAL}^{total} < 7 \text{ GeV}$, and colored lines are mass points (1, 50, 80, 90 MeV) for $e^+e^-N\to e^+A^{\prime}N$ bremsstrahlung process.}
\label{fig:missingP_brem}
\end{figure}

As discussed in \Cref{sec:bkgProduction}, the incident positron can produce multi-track backgrounds when secondary charged particles (e.g. $e^{+}e^{-}$ pairs or hadrons) are in the Tracking Detector region. By requiring exactly one tagged track (incident positron) for all three signal process and one recoil track for bremsstrahlung emission process, or zero recoil track for $e^{+}e^{-}N\to \gamma A^\prime N$ and $e^{+}e^{-}N\to A^\prime N$ annihilation processes, we suppress multi-track backgrounds while retaining signal events where the dark photon escapes undetected. For bremsstrahlung emission process $e^+N\to e^+ A^\prime N$,
the missing momentum (difference of the momentum of the tagged track and the recoil track) distribution is shown in \Cref{fig:missingP_brem}, and the cut on missing momentum $P_{tag}-P_{rec}>1$ GeV is applied during sample production.

\begin{figure}[htbp]
\centering
\includegraphics[width=.45\textwidth]{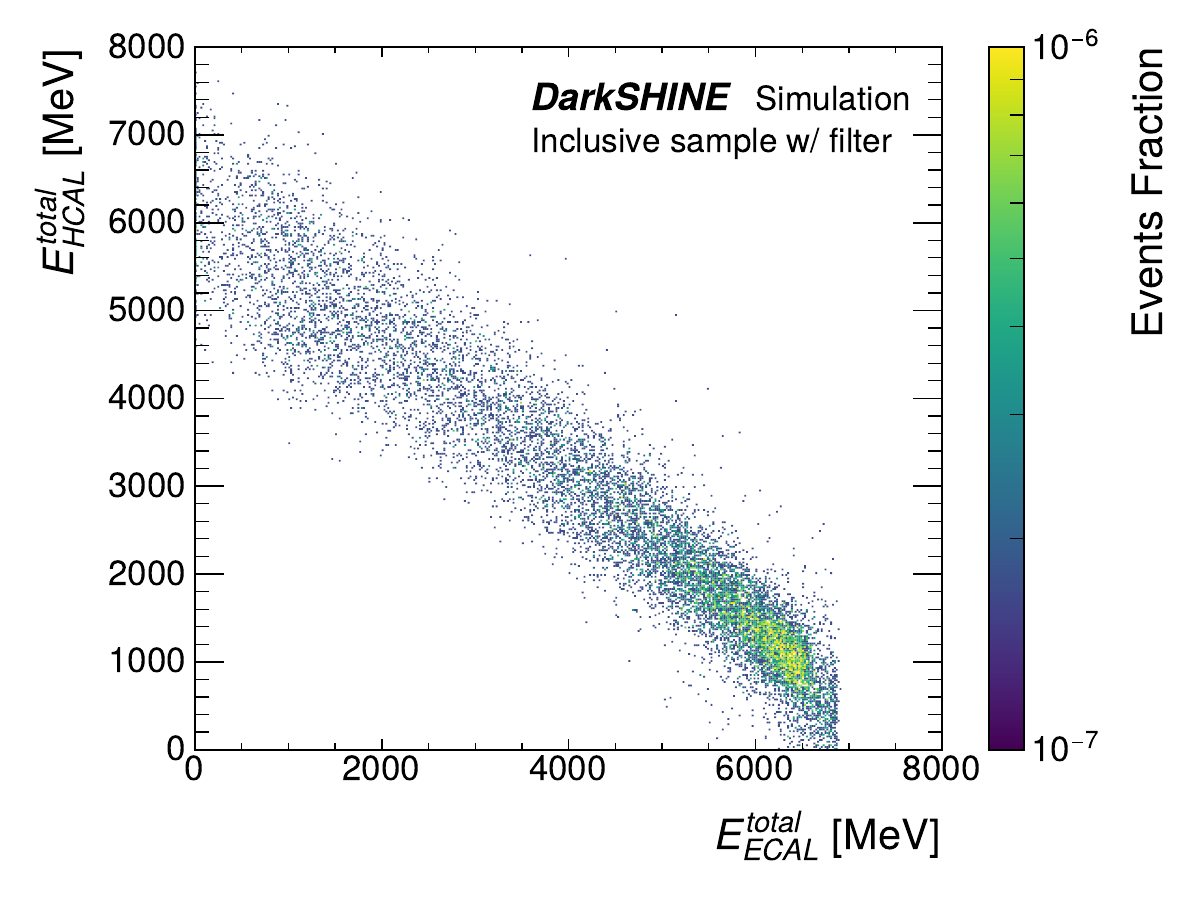}
\qquad
\includegraphics[width=.45\textwidth]{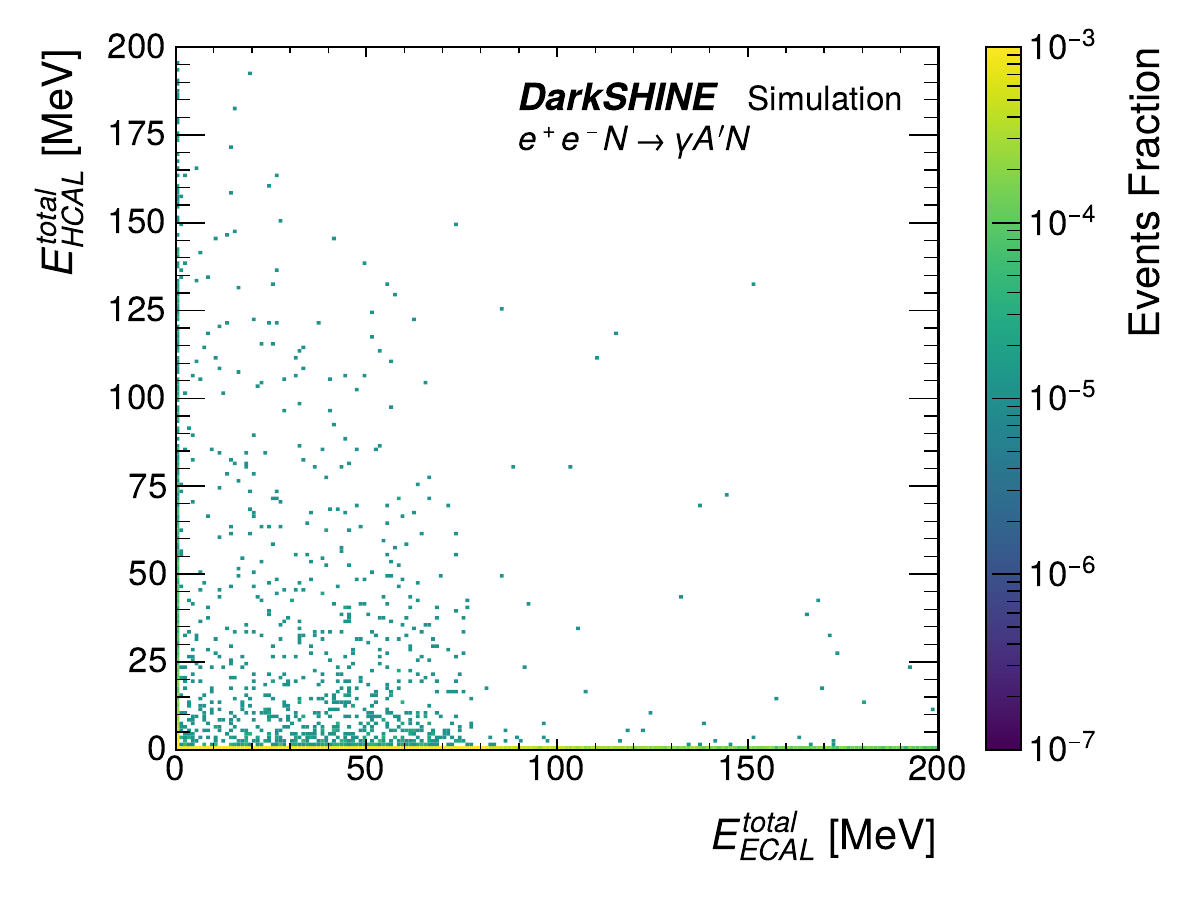}
\caption{The 2D  distribution of the ECAL and HCAL energy for the inclusive background and signal samples, respectively. The upper bound corresponds to the incident beam energy 8 GeV. }
\label{fig:ECAL_HCAL}
\end{figure}

\begin{figure}[htbp]
\centering
\includegraphics[width=.45\textwidth]{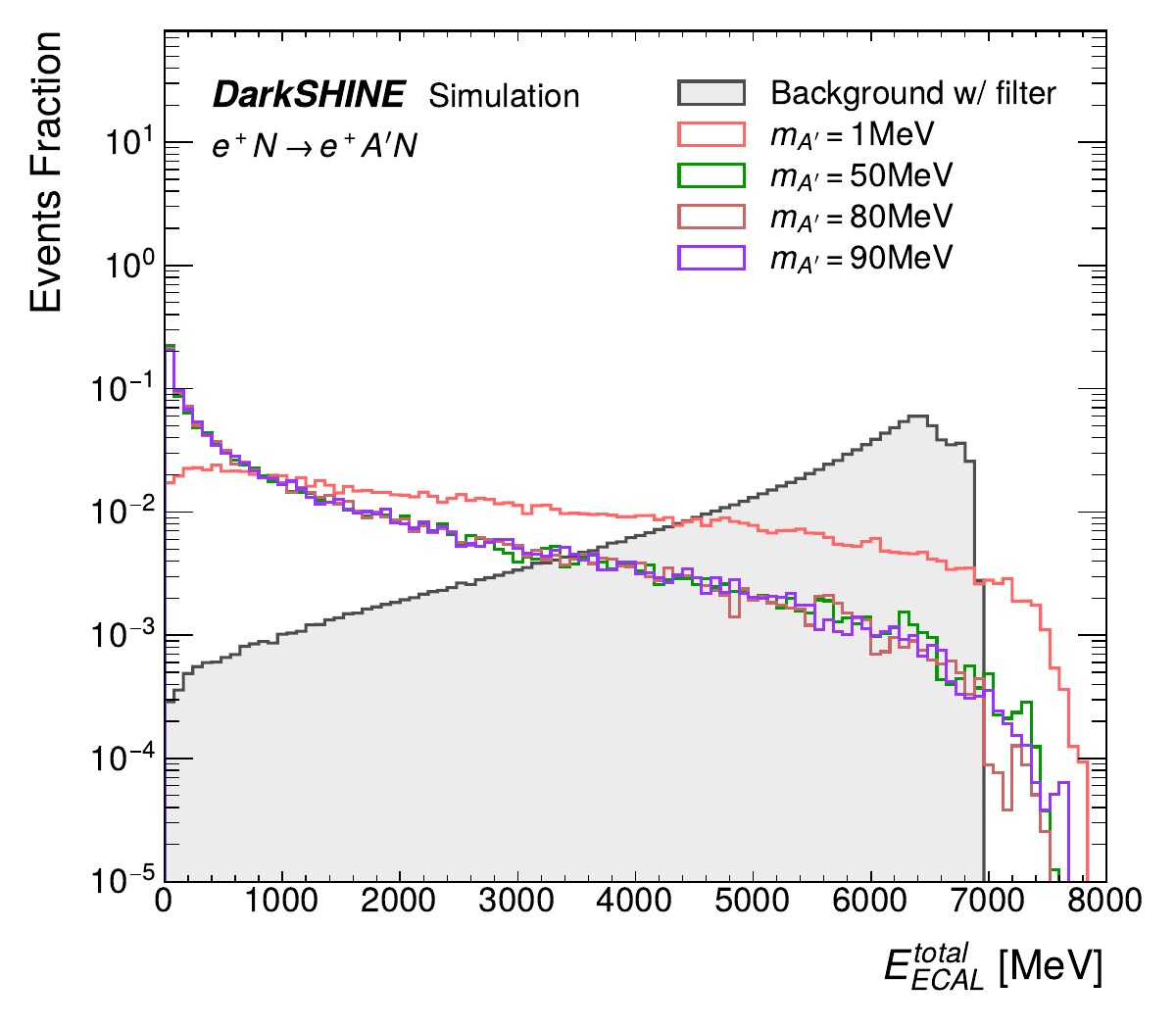}
\qquad
\includegraphics[width=.45\textwidth]{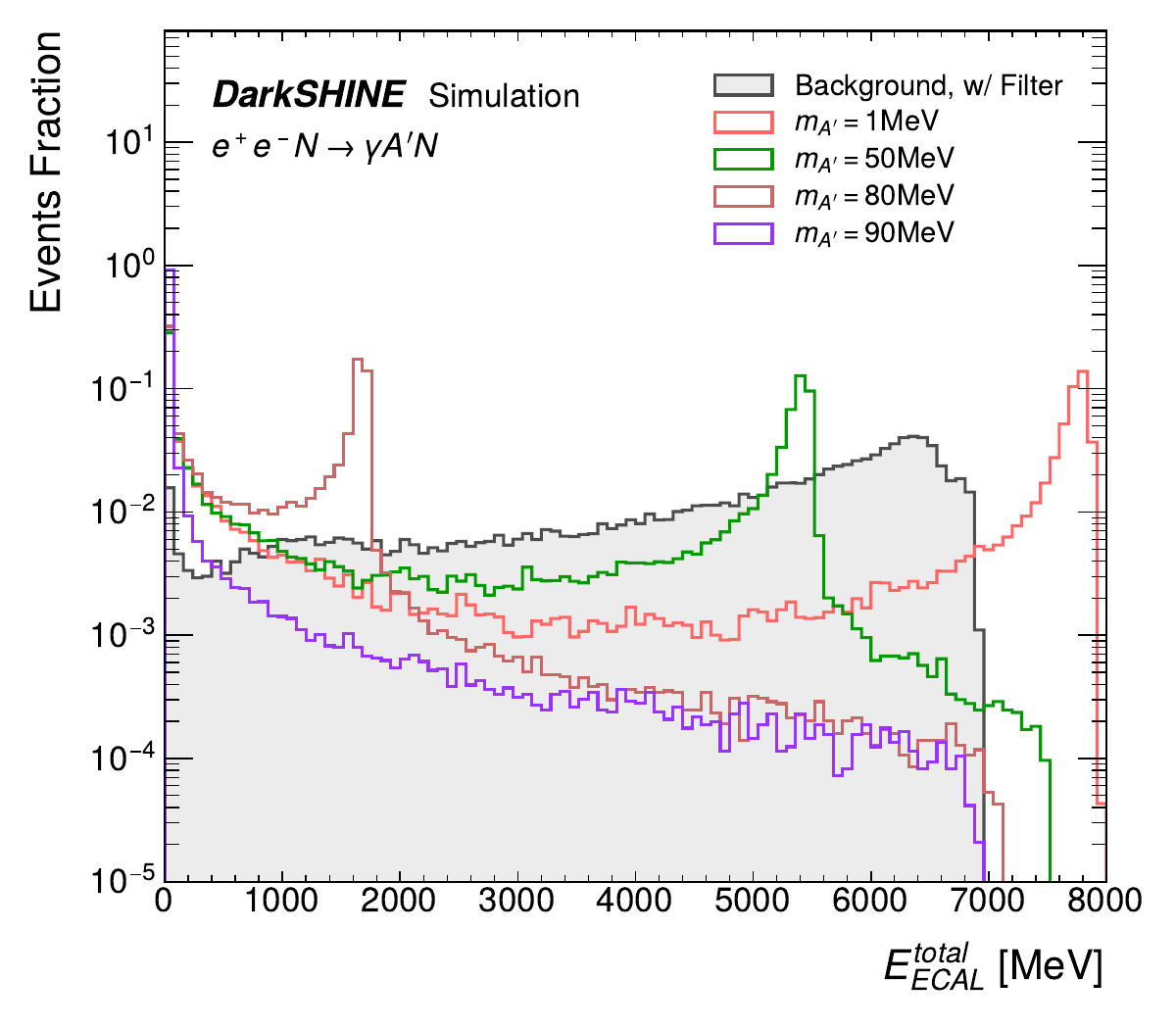}
\caption{Distributions of the total energy deposited in the electromagnetic calorimeter ($E_{\mathrm{ECAL}}^{\mathrm{tot}}$) for the inclusive background and the DarkSHINE signal processes. Left: The dark bremsstrahlung process ($e^+ N \to e^+ A^{\prime}N$), where the signal distribution shift towards lower energies. This characteristic shape arises from the missing energies carries by $A^{\prime}$. Right: The annihilation process ($e^+e^-N\to\gamma A^{\prime}N$), where the visible energy is dominated by the final-state $\gamma$. The peak structures come from the restricted kinematics of the $\gamma$ emission in this two--body-like final state, where peak positions are dependent on $m_{A^{\prime}}$. A production-level energy cut of $E_{\mathrm{ECAL}}^{\mathrm{tot}}<7\ \mathrm{GeV}$ is applied to the inclusive background (gray shaded histogram).}
\label{fig:ECAL_E_total}
\end{figure}

\begin{figure}[htbp]
\centering
\includegraphics[width=.45\textwidth]{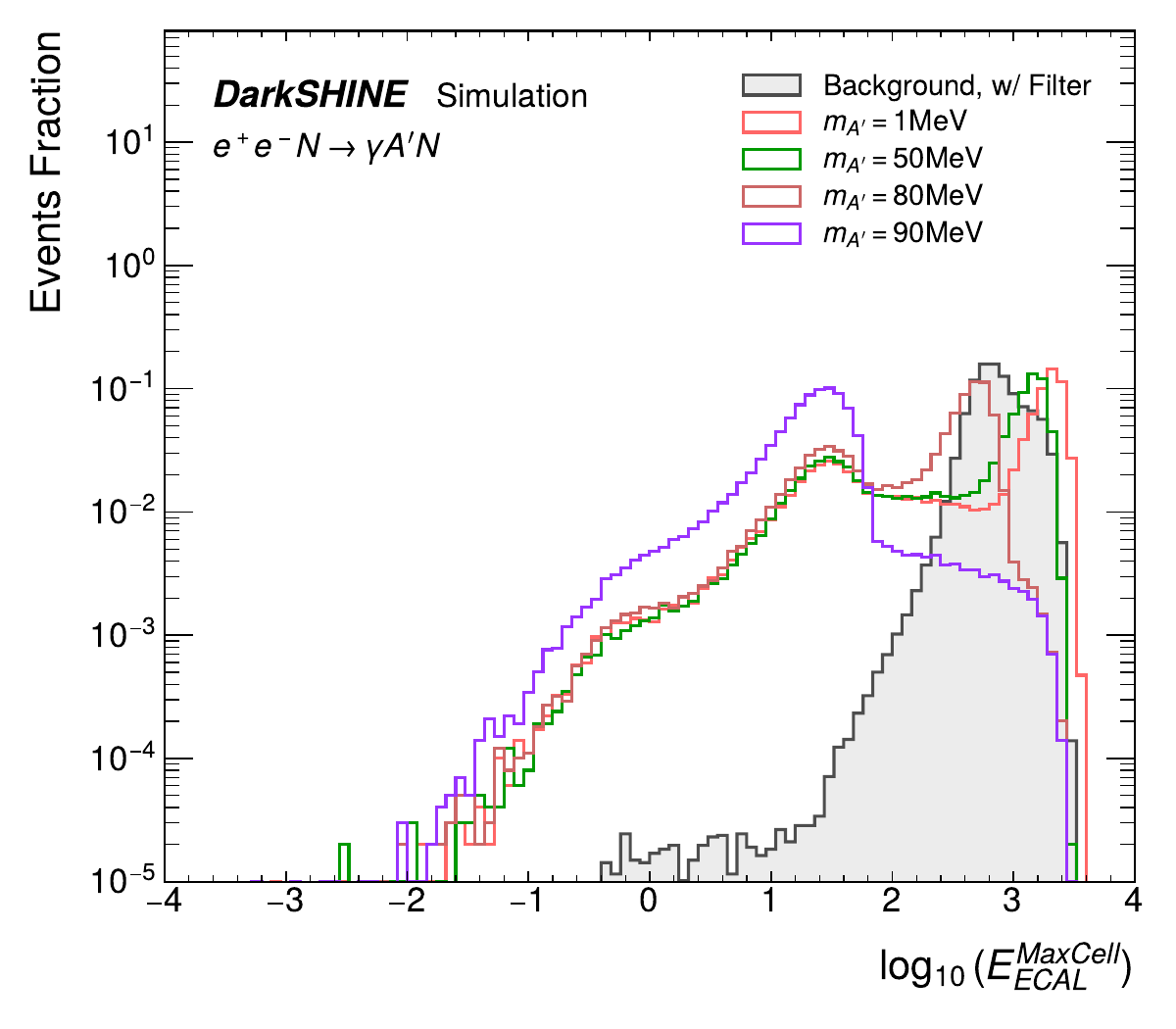}
\qquad
\includegraphics[width=.45\textwidth]{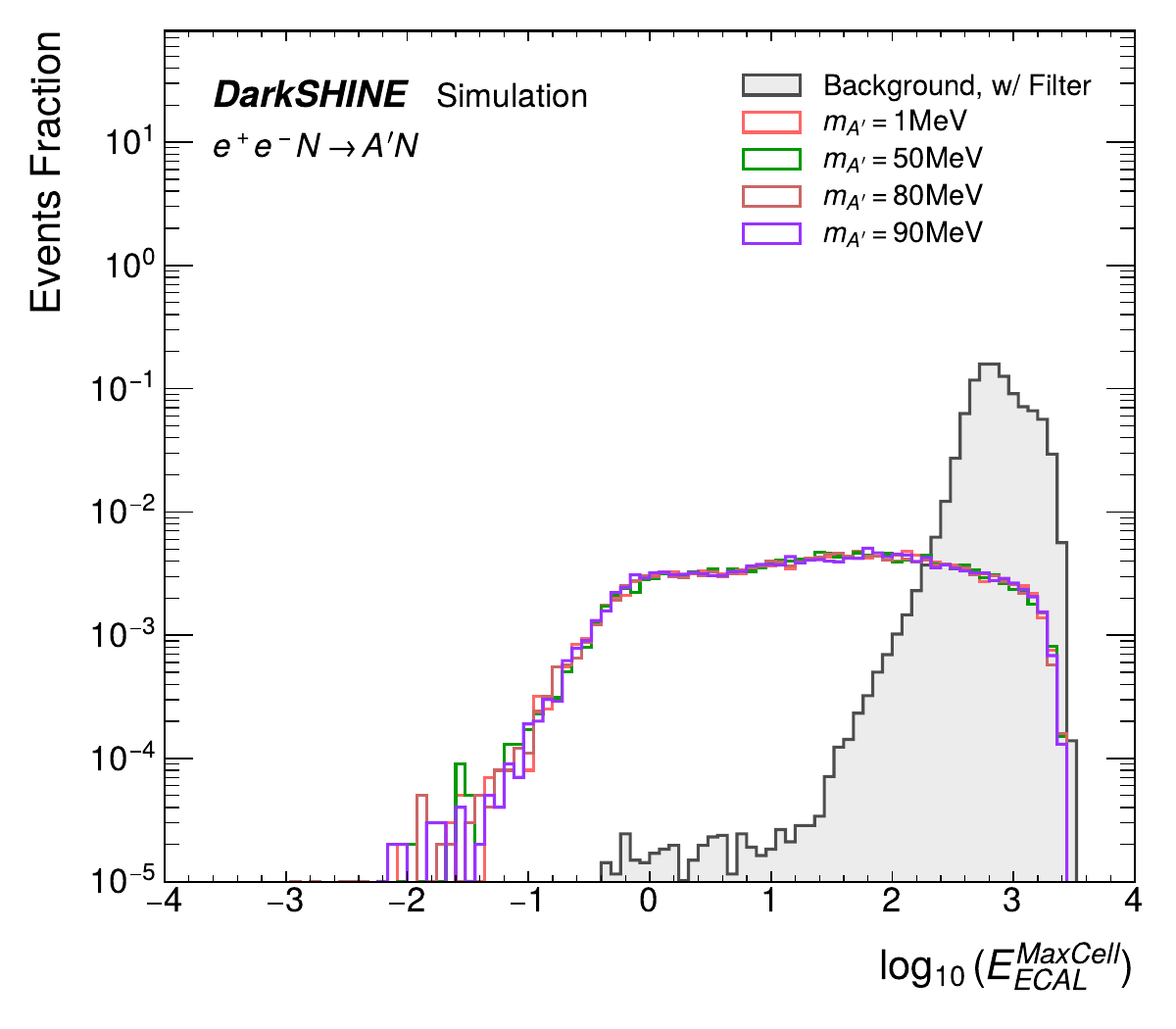}
\caption{Distributions of the logarithm of the maximum single-cell energy deposited in the ECAL for the inclusive background and dark annihilation signals. Left: The non-resonant annihilation process ($e^+e^-N \to \gamma A^{\prime}N$). The spectral shape strongly depends on the dark photon mass $m_{A^{\prime}}$, as the energetic final-state $\gamma$ governs the maximum local energy deposition in the calorimeter cells. Right: The resonant annihilation process ($e^+e^-N \to A^{\prime}N$). The signal distributions are virtually identical across different $m_{A^{\prime}}$ and are concentrated in the low-energy regime. This degeneracy arises because the $A^{\prime}$ is produced on-shell and invisibly carries away the full available center-of-mass energy, leaving only minimal energy traces (such as soft initial $e^-$ scattering in tracking detector material) to be recorded by the ECAL.}
\label{fig:ECAL_E_MaxCell}
\end{figure}

For the observables provided by the calorimeter system, \Cref{fig:ECAL_E_total} shows the total energy reconstructed in ECAL for signal ($\gamma A^{\prime}$ and $A^{\prime}$ final states) and background. In addition, the 2D distributions of the total ECAL and HCAL energy after the tracking selection criteria for the inclusive background sample and the 90 MeV non-resonant signal sample as an example shown in \Cref{fig:ECAL_HCAL}. The recoil positron of bremsstrahlung emission process and the $\gamma$ of non-resonant annihilation process can produce shower in the calorimeter. As illustrated in \Cref{fig:crossSection}, we noticed that contributions from non-resonant annihilation can be sizable at $m_{A^{\prime}}\sim 90$ MeV, in order to retain this candidate events $E_{ECAL}^{total}<2$ GeV were requested. For the bremsstrahlung emission process, we used the previously optimized value of the ECAL energy cut of 2.5 GeV. However, there is almost no energy deposition in the calorimeter since the $A^{\prime}$ decays to dark matter for $e^+e^-N\to A^{\prime}N$ annihilation process. \Cref{fig:ECAL_E_MaxCell} shows the distribution of the maximum cell energy in ECAL for the non-resonant annihilation and resonant annihilation samples, respectively. Compared with background, the maximum cell energy cut in ECAL, $E_{ECAL}^{MaxCell} \textless 1$ MeV, was applied to select resonant annihilation events, while $E_{ECAL}^{MaxCell} \geq 1$ MeV was required to select non-resonant annihilation events. On the other hand, Some events from background processes can also deposit their energy in the calorimeter detectors. For instance, muons from $\gamma \to \mu^+\mu^-$ conversions and hadronic showers from photonuclear interactions penetrate the ECAL and deposit energy in the HCAL. These background events can be effectively rejected by require the total energy and the maximum cell energy in HCAL while maintaining a high signal efficiency as discussed in Ref~\cite{DarkSHINE:2022mak}. 

\begin{figure}[htbp]
\centering
\includegraphics[width=.32\textwidth]{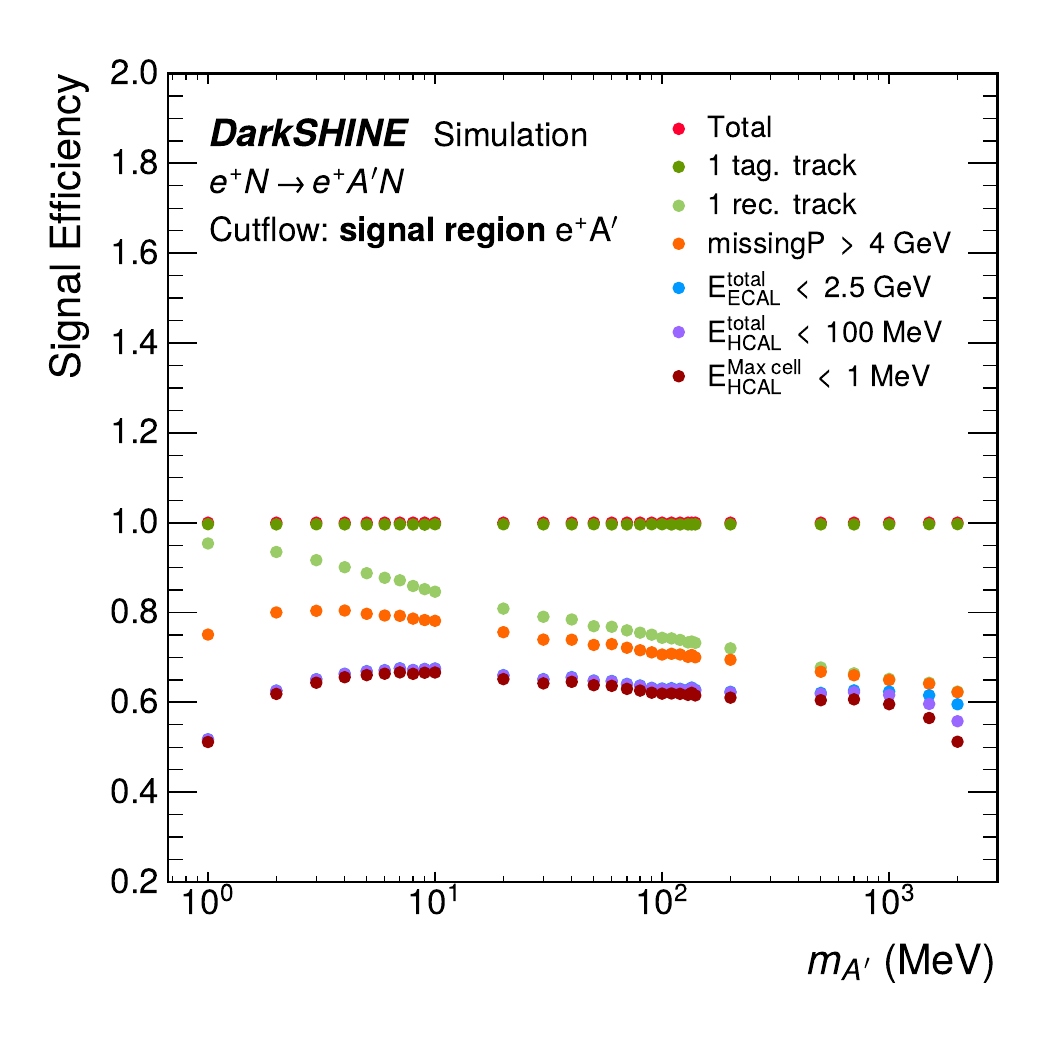}
\includegraphics[width=.32\textwidth]{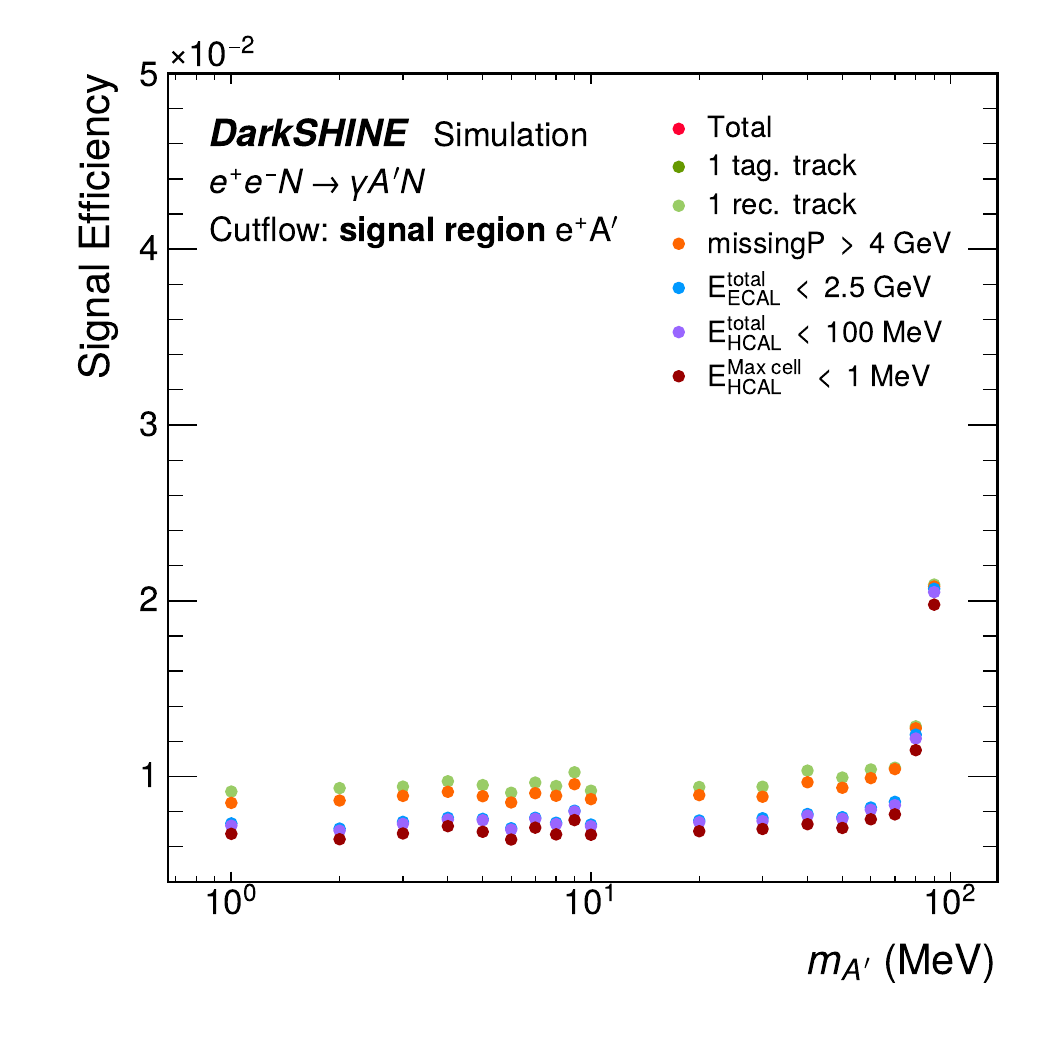}
\includegraphics[width=.32\textwidth]{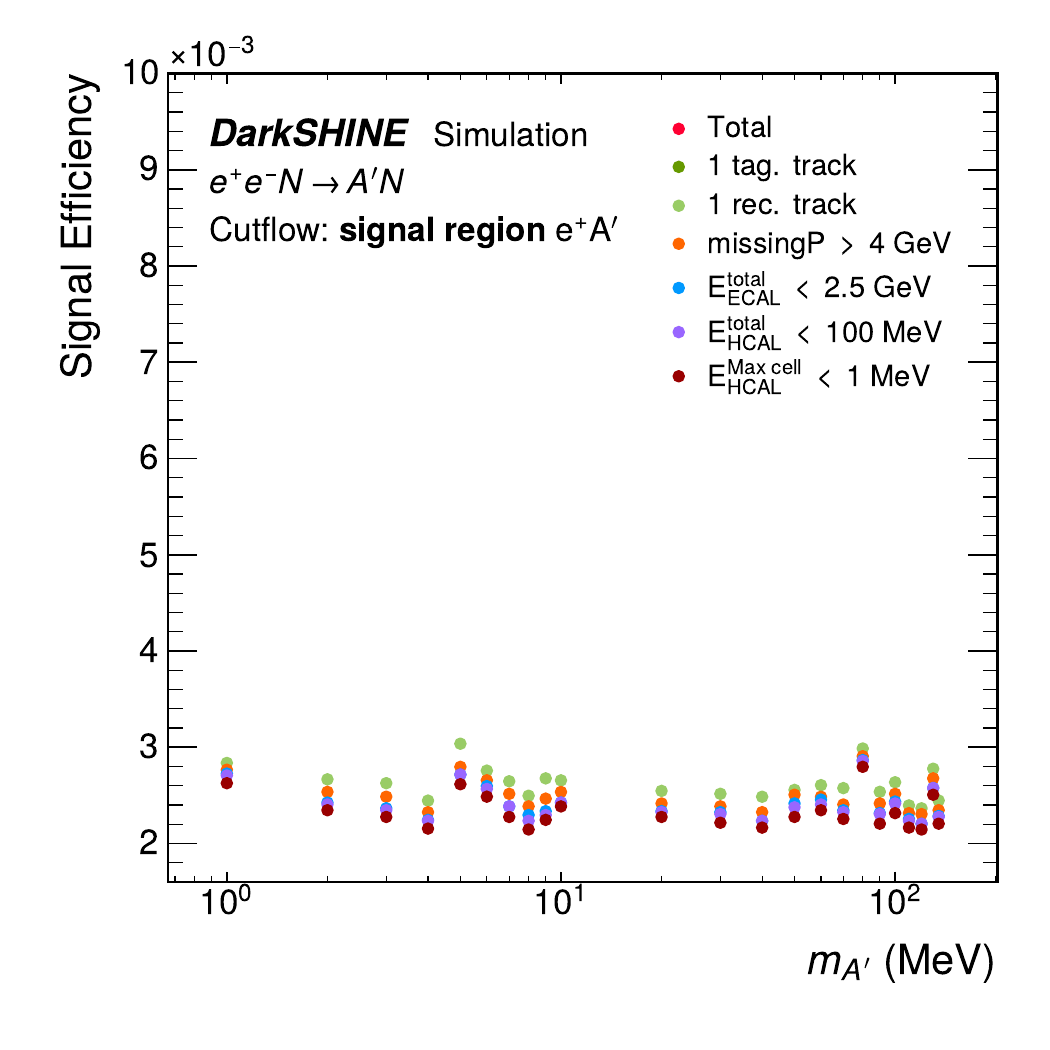}
\caption{Signal efficiencies as a function of dark photon mass ($m_{A^\prime}$) for the bremsstrahlung signal region ($e^{+}+E_{\mathrm{miss}}$). The left panel shows the efficiency for the primary target process ($e^+N \to e^+A^\prime N$). The middle and right panels illustrate the leakage of secondary signals into non-target regions. These figures shows good signal separation, with the signal region maintaining high efficiency for its target signal.}
\label{fig:eff_Brem}
\end{figure}

\begin{figure}[htbp]
\centering
\includegraphics[width=.32\textwidth]{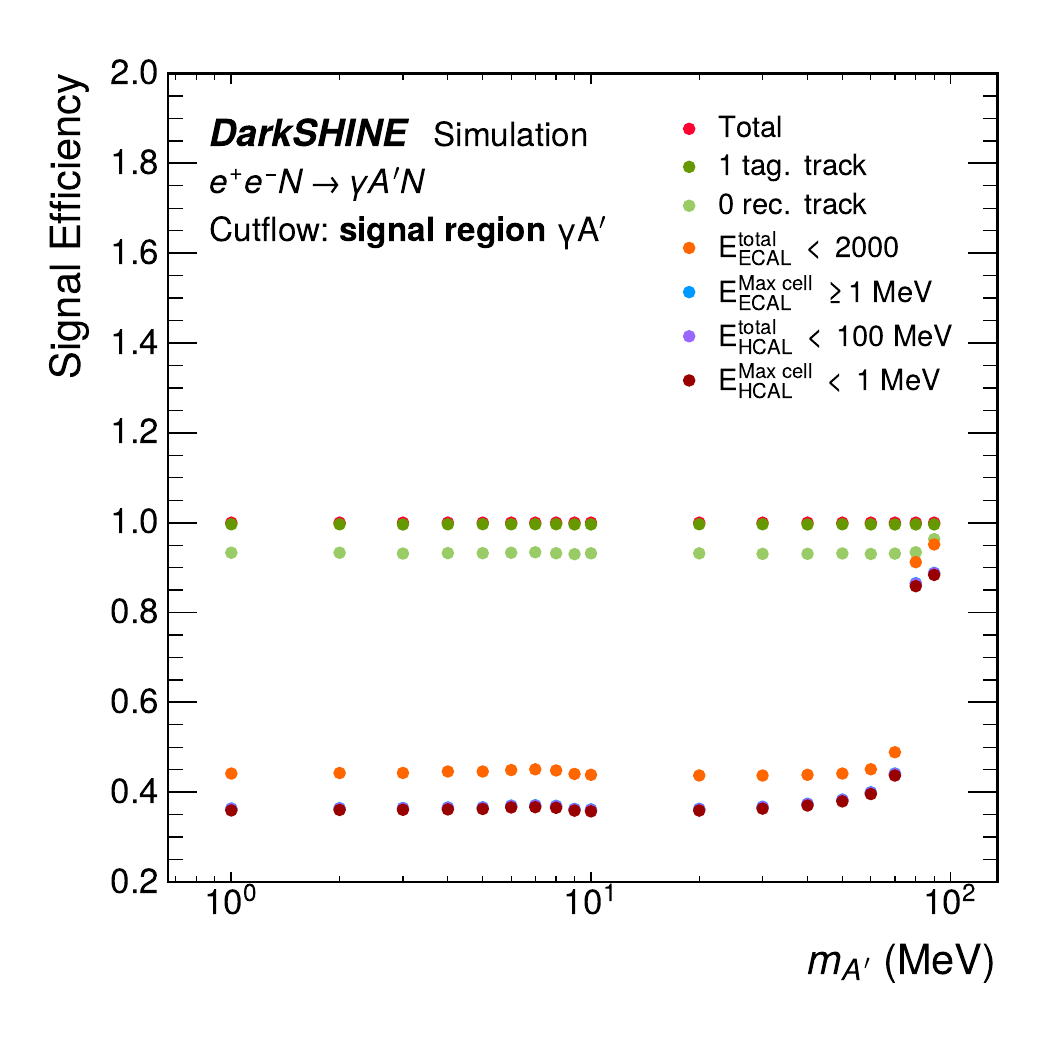}
\includegraphics[width=.32\textwidth]{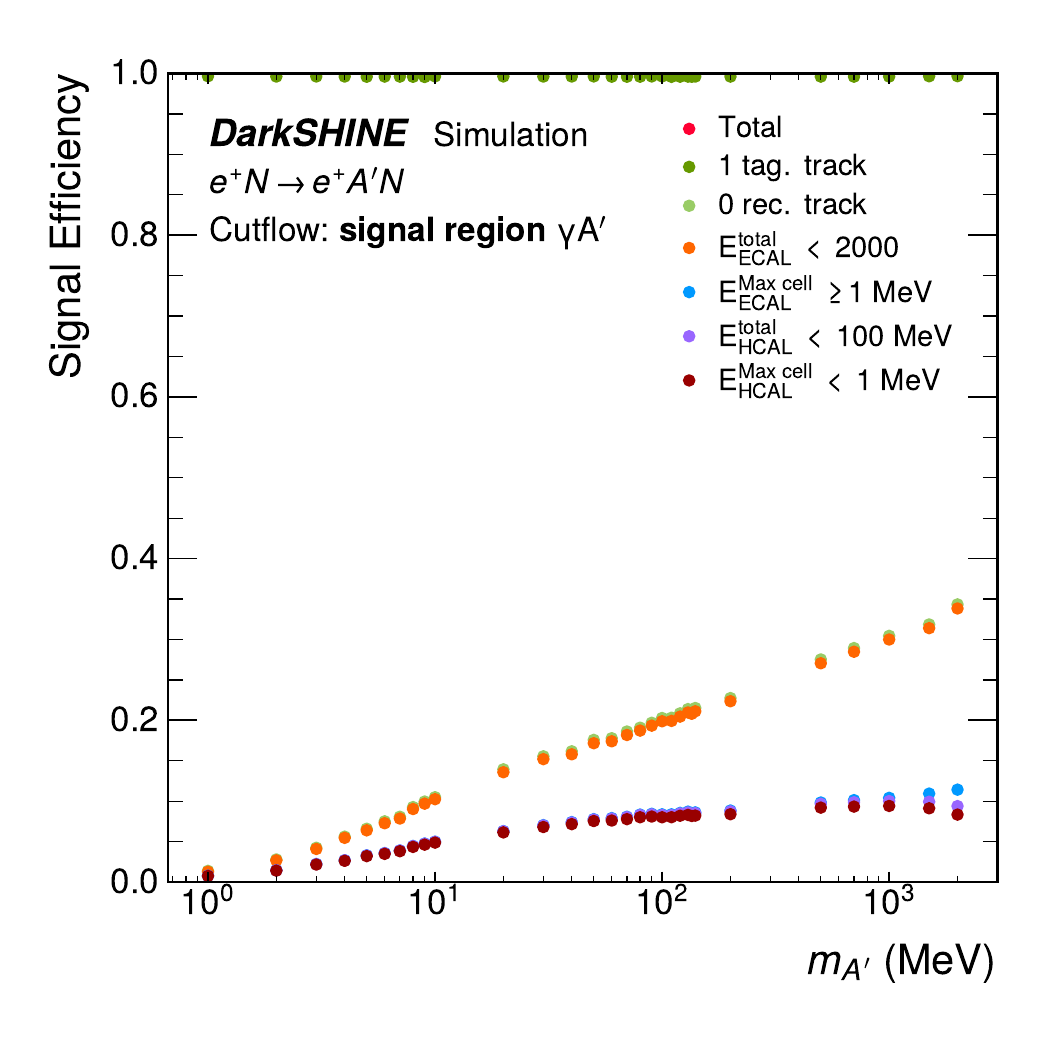}
\includegraphics[width=.32\textwidth]{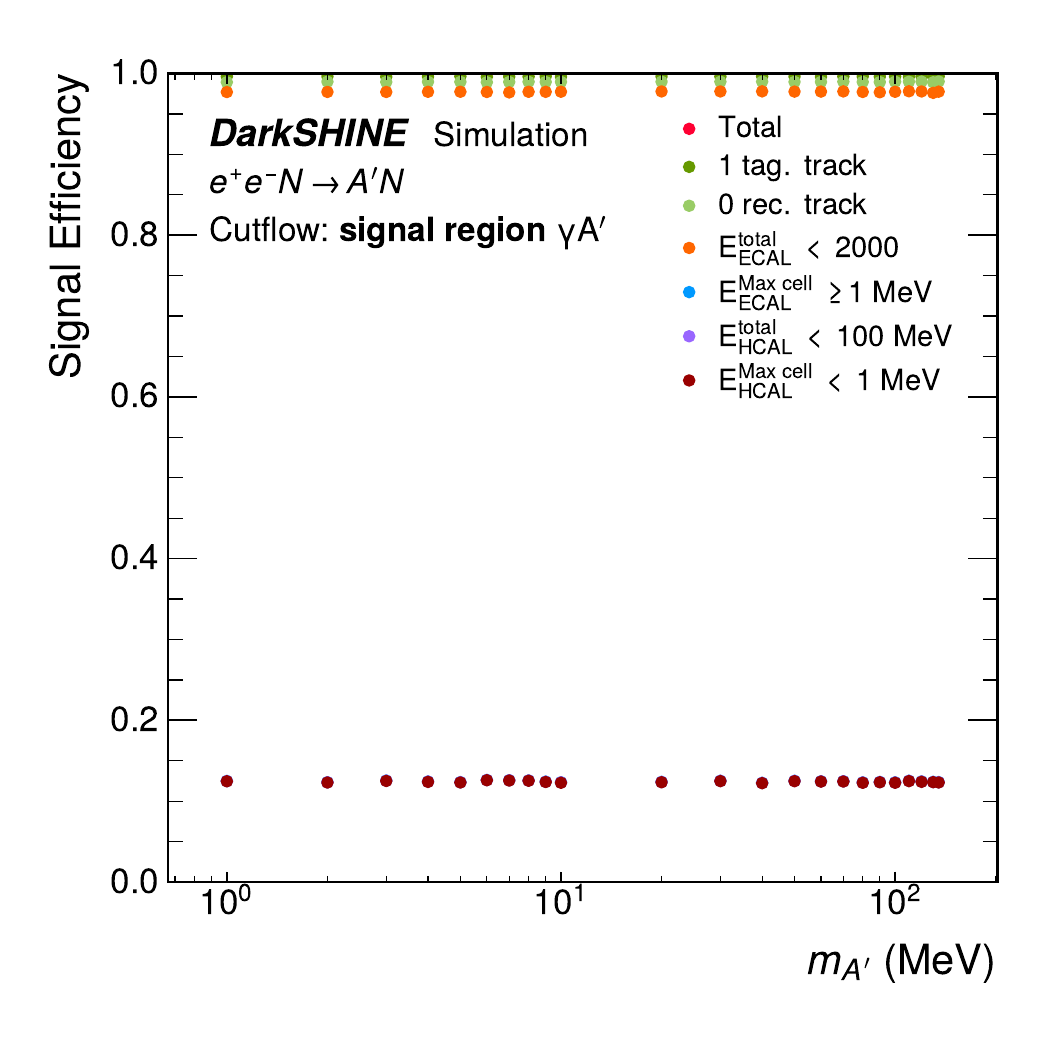}
\caption{Signal efficiencies as a function of dark photon mass ($m_{A^\prime}$) for the non-resonant signal region ($\gamma+E_{\mathrm{miss}}$).  The left panel shows the efficiency for the primary target process ( $e^+e^-N \to \gamma A^{\prime}N$ ). The middle and right panels display the secondary signals' leakage into non-target regions.}
\label{fig:eff_NR}
\end{figure}

\begin{figure}[htbp]
\centering
\includegraphics[width=.32\textwidth]{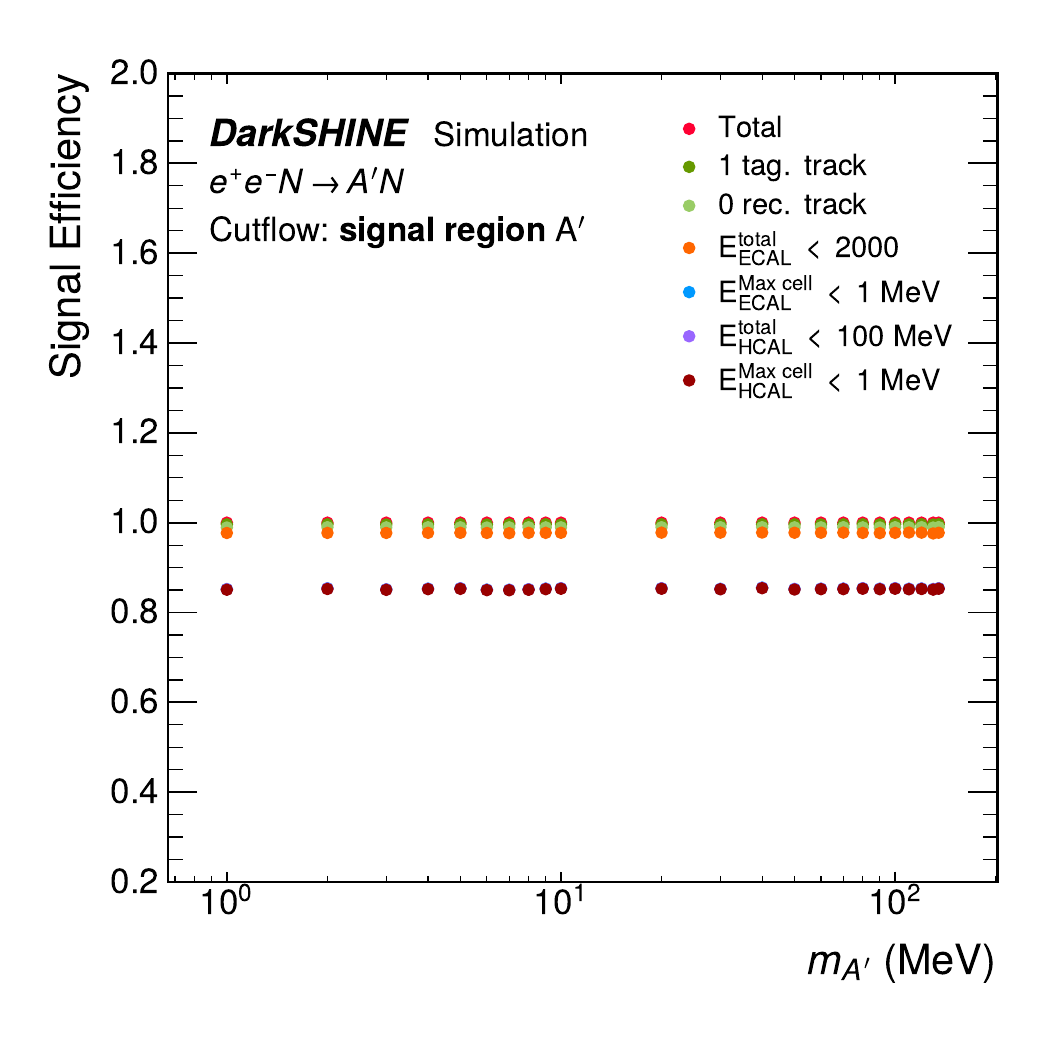}
\includegraphics[width=.32\textwidth]{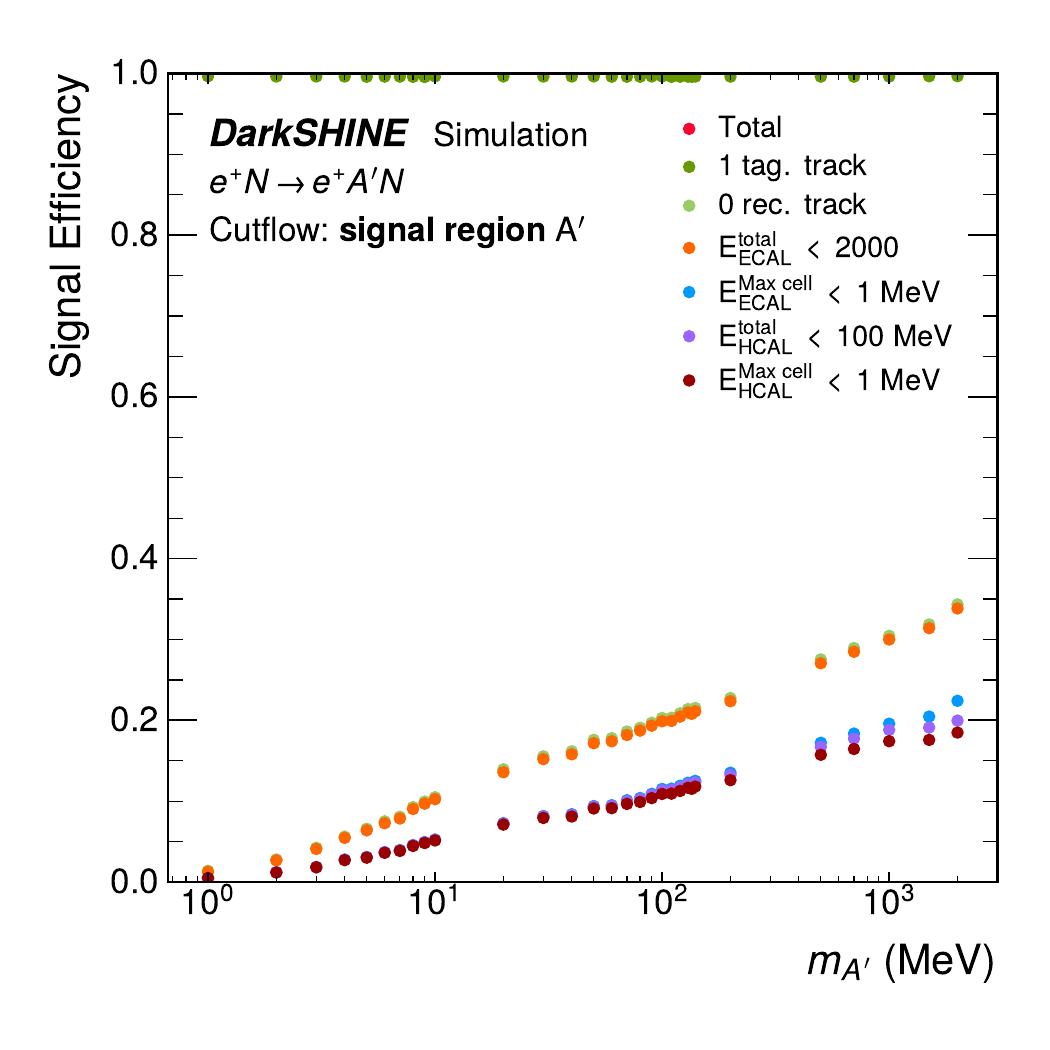}
\includegraphics[width=.32\textwidth]{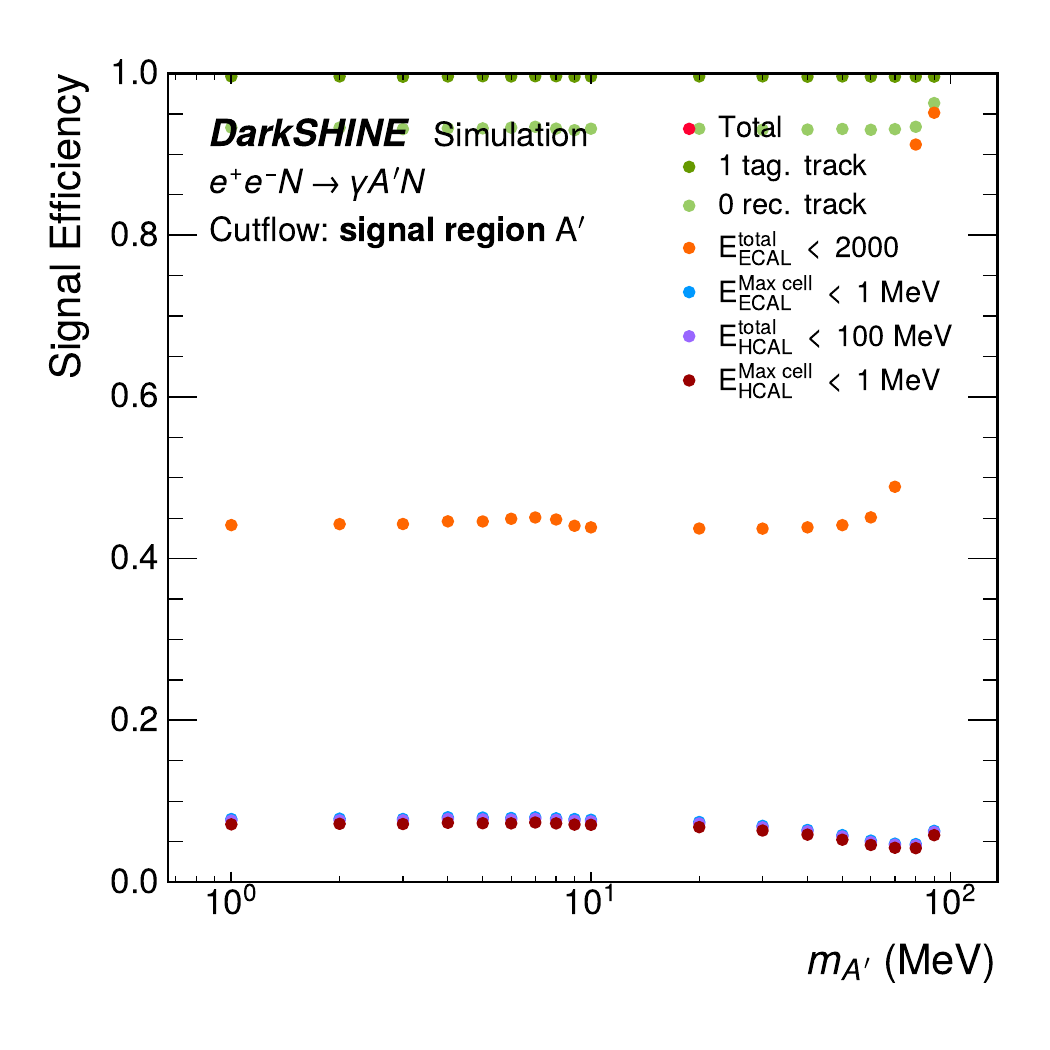}
\caption{Signal efficiencies as a function of dark photon mass ($m_{A^\prime}$) for the resonant signal region ($E_{\mathrm{miss}}$). The left panel shows the efficiency for the primary target process ($e^+e^-N \to A^{\prime}N$). The middle and right panels display the secondary signals' leakage into non-target regions.}
\label{fig:eff_Res}
\end{figure}

In consequence, the signal efficiencies for the three signal processes (bremsstrahlung emission $e^+e^-N\to e^+A^{\prime}N$ , t-channel production $e^+e^-N\to \gamma A^{\prime} N$, s-channel production $e^+e^-N \to A^{\prime} N$) in each of the three signal regions ($e^++E_{\mathrm{miss}}$, $\gamma+E_{\mathrm{miss}}$, $E_{\mathrm{miss}}$) were evaluated using signal samples simulated with DarkSHINE software as shown in \Cref{fig:eff_Brem}--\ref{fig:eff_Res}. The 9 subfigures demonstrate excellent signal separation capability, with each signal region maintaining higher efficiency for its target signal while keeping cross-contamination below $1\%$. The left panels in each figure show the kinematic distributions for the primary signal, while the middle and right panels display the secondary signals' leakage into non-target regions. 

\begin{table}[htbp]
\centering
\scriptsize
\caption{Bremsstrahlung signal region ($e^{+}+E_{\mathrm{miss}}$) cutflow for inclusive background and rare processes samples. The number of events remaining after each cut is listed in the table.}
\label{tab:cutflow_Brem}
\resizebox{\textwidth}{!}{
\begin{tabular}{lccccccccc}
\hline
Sample & POT & Generated & Filter & $N_{trk}^{tag}=1$ & $N_{trk}^{rec}=1$ & \makecell{$\Delta P >$ \\ 4 GeV} & \makecell{$E_{HCAL}^{total}$ \\ $< 0.1$ MeV} & \makecell{$E_{HCAL}^{Max}$ \\ $<1$ MeV} & \makecell{$E_{ECAL}^{total}$ \\ $< 2.5$ GeV} \\
\hline
PN ECAL & $1.48\times10^{11}$ & $1\times10^8$ & 5829159 & 5794358 & 4641895 & 3114535 & 272207 & 88871 & 0 \\
PN Target & $4.99\times10^{12}$ & $1\times10^8$ & 5623324 & 5386830 & 455540 & 429175 & 11831 & 2459 & 0 \\
EN Target & $2.78\times10^{12}$ & $1\times10^8$ & 1385618 & 1326519 & 117549 & 114845 & 977 & 206 & 0 \\
Annihil Target & $3.38\times 10^{12}$ & $1\times10^8$ & 289221 & 288943 & 5595 & 5466 & 201 & 60 & 0 \\
GMM ECAL & $2.02\times10^{13}$ & $1\times10^8$ & 9333419 & 9331661 & 8842313 & 6123627 & 6 & 6 & 0 \\
EN ECAL & $3.23\times10^{11}$ & $1\times10^8$ & 124934 & 124241 & 99812 & 58 & 5 & 2 & 0 \\
GMM Target & $4.73\times10^{13}$ & $1\times10^7$ & 663639 & 663576 & 20 & 0 & 0 & 0 & 0 \\
Inclusive & $2.5\times10^9$ & $2.5\times10^9$ & 1472414 & 1470635 & 1359999 & 570317 & 7175 & 2077 & 0 \\
\hline
\end{tabular}
}
\end{table}

\begin{table}[htbp]
\centering
\scriptsize
\caption{Signal region ($\gamma+E_{\mathrm{miss}}$) cutflow for inclusive background and rare processes samples. The number of events remaining after each cut is listed in the table.}
\label{tab:cutflow_NR}
\resizebox{\textwidth}{!}{
\begin{tabular}{lccccccccc}
\hline
Sample & POT & Generated & Filter & $N_{trk}^{tag}=1$ & $N_{trk}^{rec}=0$ & \makecell{$E_{ECAL}^{MaxCell}$\\$\geq1$ MeV} & \makecell{$E_{HCAL}^{total}$\\$<0.1$ GeV} & \makecell{$E_{HCAL}^{MaxCell}$\\$<1$ MeV} & \makecell{$E_{ECAL}^{total}$\\$<2$ GeV} \\
\hline
PN ECAL & $1.48\times10^{11}$ & $1\times 10^8$ & 5829159 & 5794358 & 138513 & 138513 & 8051 & 2552 & 0 \\
Annihil Target & $3.38\times 10^{12}$ & $1\times10^8$ & 289221 & 288943 & 20649 & 20569 & 768 & 241 & 0 \\
PN Target & $4.99\times10^{12}$ & $1\times 10^8$ & 5623324 & 5386830 & 5391 & 5391 & 162 & 40 & 0 \\
EN Target & $2.78\times10^{12}$ & $1\times 10^8$ & 1385618 & 1326519 & 1633 & 1633 & 8 & 0 & 0 \\
GMM ECAL & $2.02\times10^{13}$ & $1\times 10^8$ & 9333419 & 9331661 & 67950 & 67950 & 0 & 0 & 0 \\
EN ECAL & $3.23\times10^{11}$ & $1\times 10^8$ & 124934 & 124241 & 0 & 0 & 0 & 0 & 0 \\
GMM Target & $4.73\times10^{13}$ & $1\times 10^7$ & 663639 & 663576 & 0 & 0 & 0 & 0 & 0 \\
Inclusive & $2.5\times 10^9$ & $2.5\times 10^9$ & 1472414 & 1470635 & 9240 & 9208 & 178 & 53 & 0 \\
\hline
\end{tabular}
}
\end{table}

\begin{table}[htbp]
\centering
\scriptsize
\caption{Signal region ($E_{\mathrm{miss}}$) cutflow for inclusive background and rare processes samples. The number of events remaining after each cut is listed in the table.}
\label{tab:cutflow_Res}
\resizebox{\textwidth}{!}{
\begin{tabular}{lccccccccc}
\hline
Sample & POT & Generated & Filter & $N_{trk}^{tag}=1$ & $N_{trk}^{rec}=0$ & \makecell{$E_{ECAL}^{total}$\\$<2.0$ GeV} & \makecell{$E_{HCAL}^{total}$\\$<0.1$ GeV} & \makecell{$E_{HCAL}^{MaxCell}$\\$<1$ MeV} & \makecell{$E_{ECAL}^{MaxCell}$\\$<1$ MeV} \\
\hline
PN ECAL & $1.48\times10^{11}$ & $1\times 10^8$ & 5829159 & 5794358 & 138513 & 120147 & 3922 & 1096 & 0 \\
PN Target & $4.99\times10^{12}$ & $1\times 10^8$ & 5623324 & 5386830 & 5391 & 4176 & 66 & 14 & 0 \\
EN Target & $2.78\times10^{12}$ & $1\times 10^8$ & 1385618 & 1326519 & 1633 & 1554 & 2 & 0 & 0 \\
GMM ECAL & $2.02\times10^{13}$ & $1\times 10^8$ & 9333419 & 9331661 & 67950 & 67950 & 0 & 0 & 0 \\
Annihil Target & $3.38\times 10^{12}$ & $1\times10^8$ & 289221 & 288943 & 20649 & 1111 & 0 & 0 & 0 \\
GMM Target & $4.73\times10^{13}$ & $1\times 10^7$ & 663639 & 663576 & 0 & 0 & 0 & 0 & 0 \\
EN ECAL & $3.23\times10^{11}$ & $1\times 10^8$ & 124934 & 124241 & 0 & 0 & 0 & 0 & 0 \\
Inclusive & $2.5\times 10^9$ & $2.5\times 10^9$ & 1472414 & 1470635 & 9240 & 8390 & 61 & 11 & 0 \\
\hline
\end{tabular}
}
\end{table}

\section{Background estimation}
\label{sec:estimateBkg}
As illustrated in \Cref{tab:cutflow_Brem}--\ref{tab:cutflow_Res}, none of the simulated inclusive background events ($2.5 \times 10^9$) pass the event selection, similar to the situation for each rare background process. In order to estimate background events for one year run of DarkSHINE experiment with $3\times10^{14}$ POTs, an extrapolation method was used separately in three signal regions ($e^++E_{\mathrm{miss}}$, $\gamma+E_{\mathrm{miss}}$, $E_{\mathrm{miss}}$). This approach follows the previous analysis in Ref.~\cite{DarkSHINE:2022mak} which estimate background in signal region by exploit the trend of the background rejection efficiency in the side-band regions. The details of procedure from rare background processes described in \Cref{sec:estimateBkg_rare}. In addition, the same method applied with inclusive sample as a validation was discussed in \Cref{sec:estimateBkg_inclusive}.

\subsection{Extrapolation from rare processes simulation}
\label{sec:estimateBkg_rare}

\begin{figure}[htbp]
\centering
\includegraphics[width=.6\textwidth]{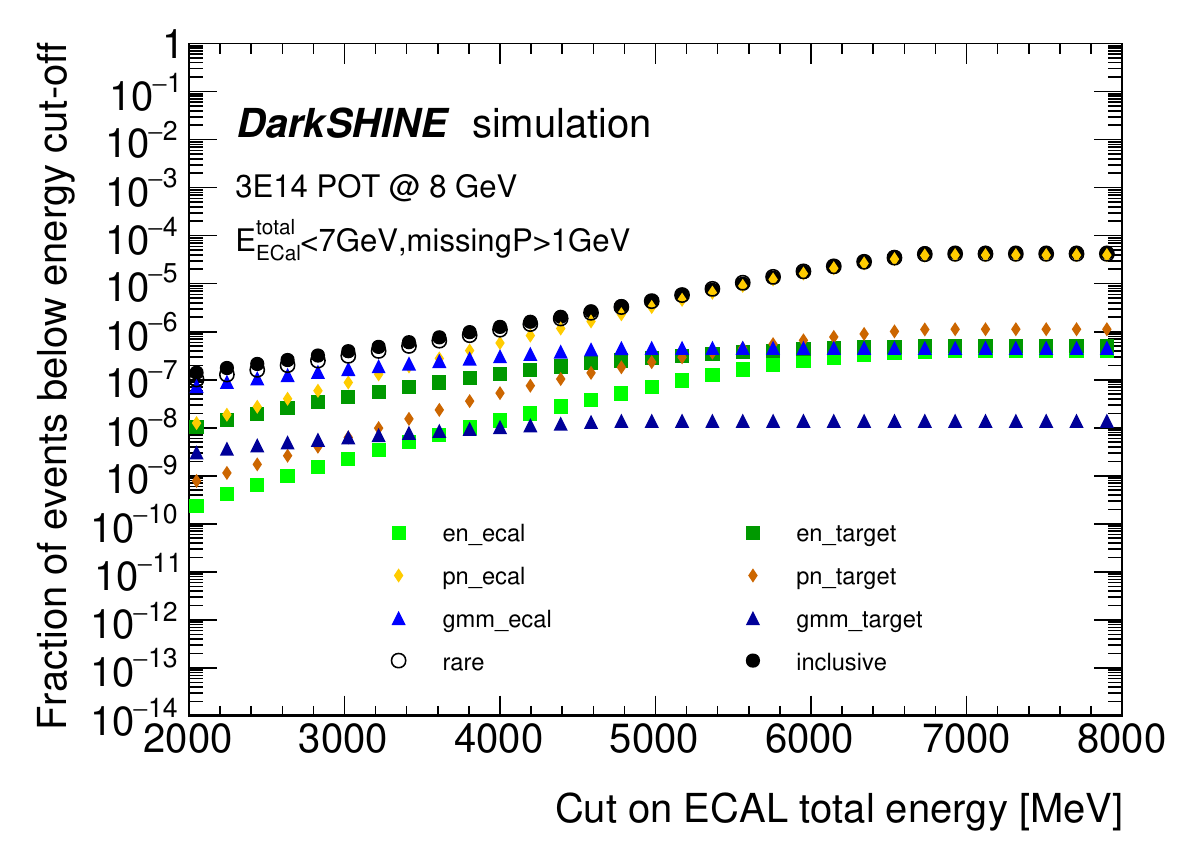}
\caption{Event ratio as a function of the cut value on ECAL energy.}
\label{fig:validation}
\end{figure}

There are six rare background processes as discussed in \Cref{sec:bkgProduction}: EN ECAL, EN Target, PN ECAL, PN Target, GMM ECAL, and GMM Target. \Cref{fig:validation} shows the fraction of background events remaining with a certain ECAL energy cut with samples listed in \Cref{tab:bkg_summary}, requiring a pre-selection ($P_{tag}-P_{rec}>1$ GeV and $E_{ECAL}^{total}<7$ GeV). In this plot, the event ratios of six rare background processes are scaled according to their branching ratios. For each rare process, the event ratio derived from the individually produced sample agrees with that obtained from the inclusive background sample. The combined contribution of these six processes is represented by gray circles (dots), obtained from the rare (inclusive) background samples. It is observed that the two summed fractions exhibit good agreement, particularly in the high-energy region where sufficient statistics are available. As we can see, the expected background yield can be derived from the corresponding event ratio by giving an ECAL energy cut threshold. For instance, if the event fraction falls below $10^{-14}$ at a cut value $x$, one can safely conclude that for a total of $10^{14}$ events, fewer than one background event will remain when applying the requirement $E_{\text{ECAL}}^{\text{total}} < x$ MeV. To estimate the background yield in each signal region as outlined in \Cref{sec:signal_region}, we must implement all corresponding selections. However, after applying all the cuts, we cannot directly determine this number due to limited statistics. Because the HCAL energy cuts are so stringent that most processes shown in \Cref{fig:validation} start to run out of statistics before the nominal ECAL energy cut threshold applied. To address this limitation, we produced dedicated rare background processes with enhanced statistics and used an extrapolation method to extend the event ratio trends into the low-energy region. The background yields in each signal region ($e^{+}+E_{\mathrm{miss}}$, $\gamma+E_{\mathrm{miss}}$, and $E_{\mathrm{miss}}$) were evaluated with the six rare background channels.

\begin{figure}[htbp]
\centering
\includegraphics[width=.4\textwidth]{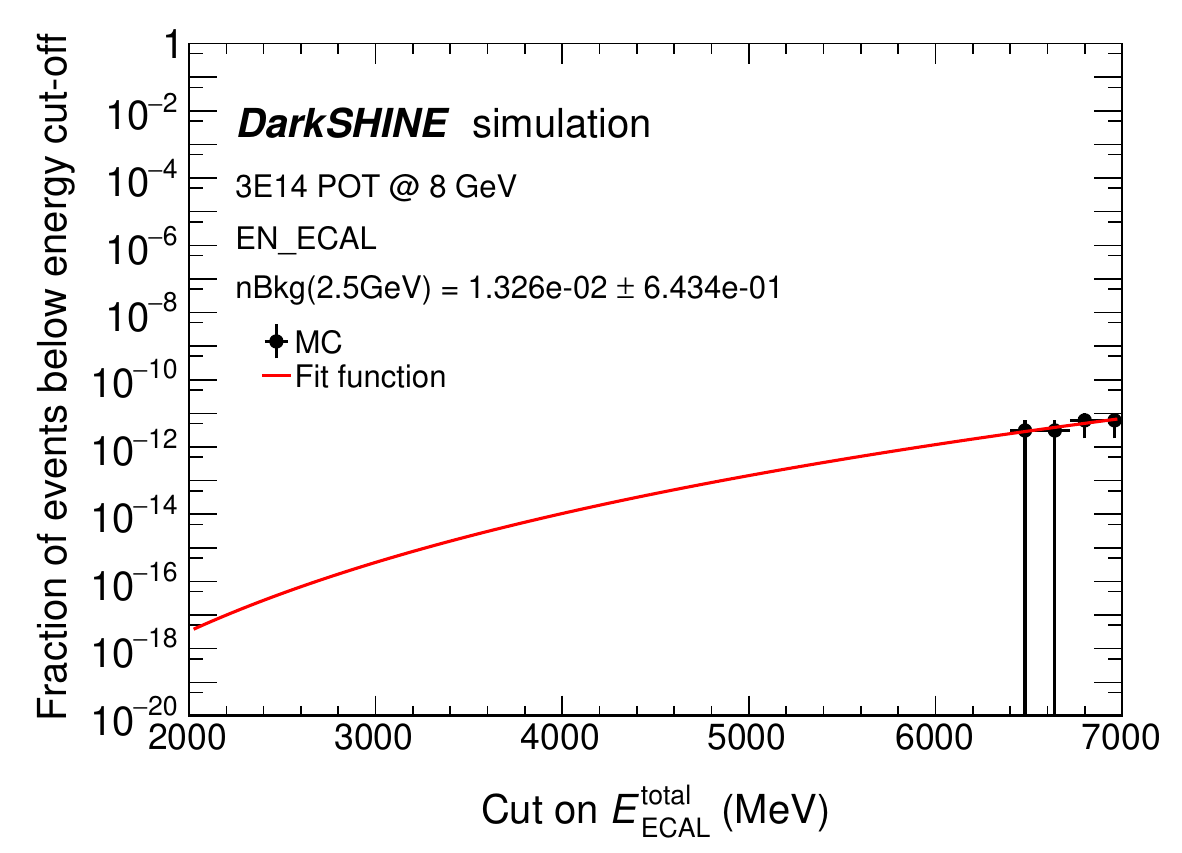}
\includegraphics[width=.4\textwidth]{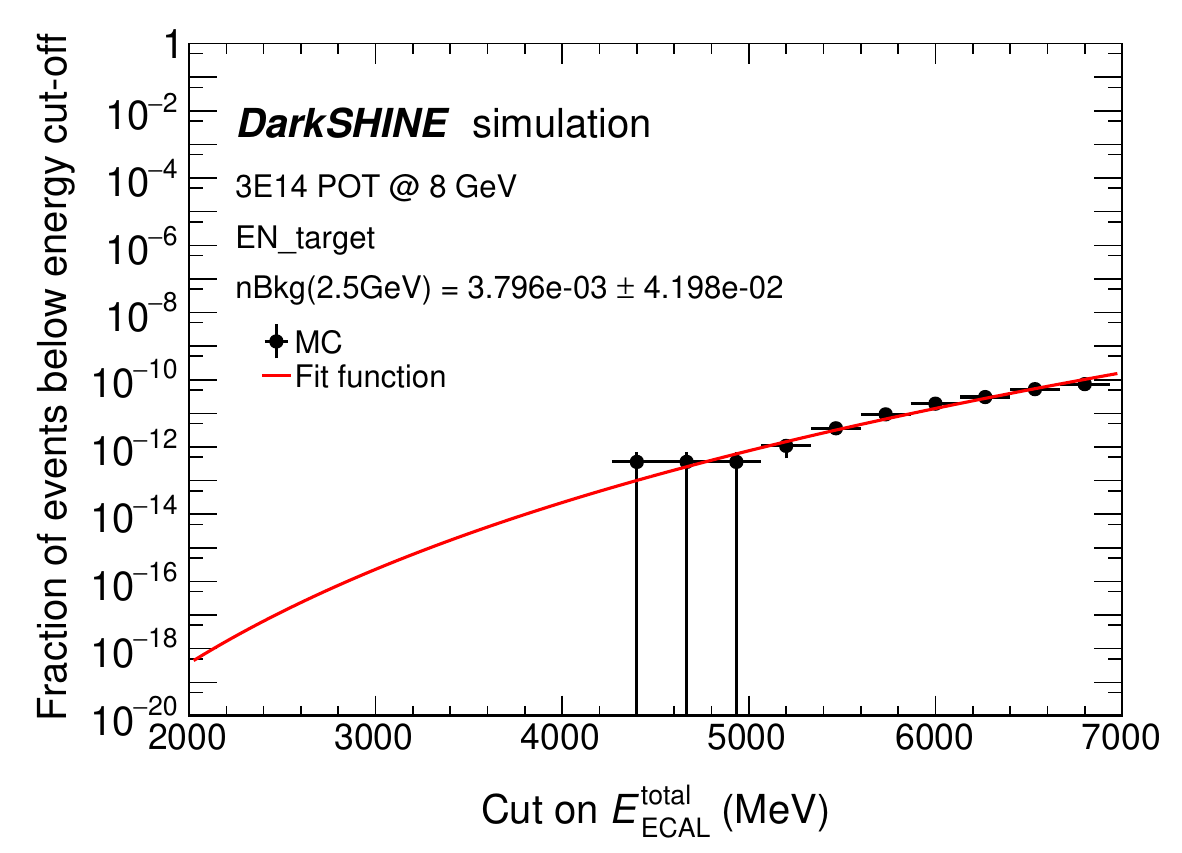}
\includegraphics[width=.4\textwidth]{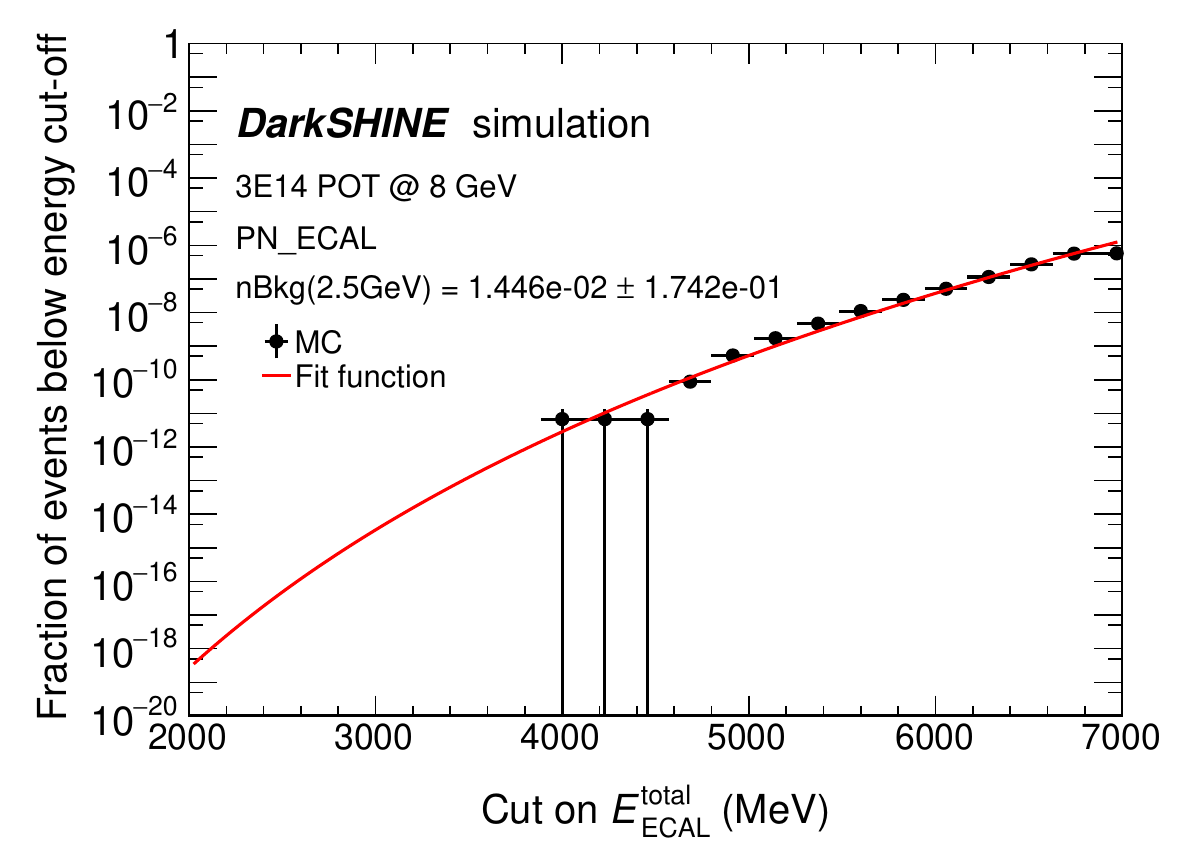}
\includegraphics[width=.4\textwidth]{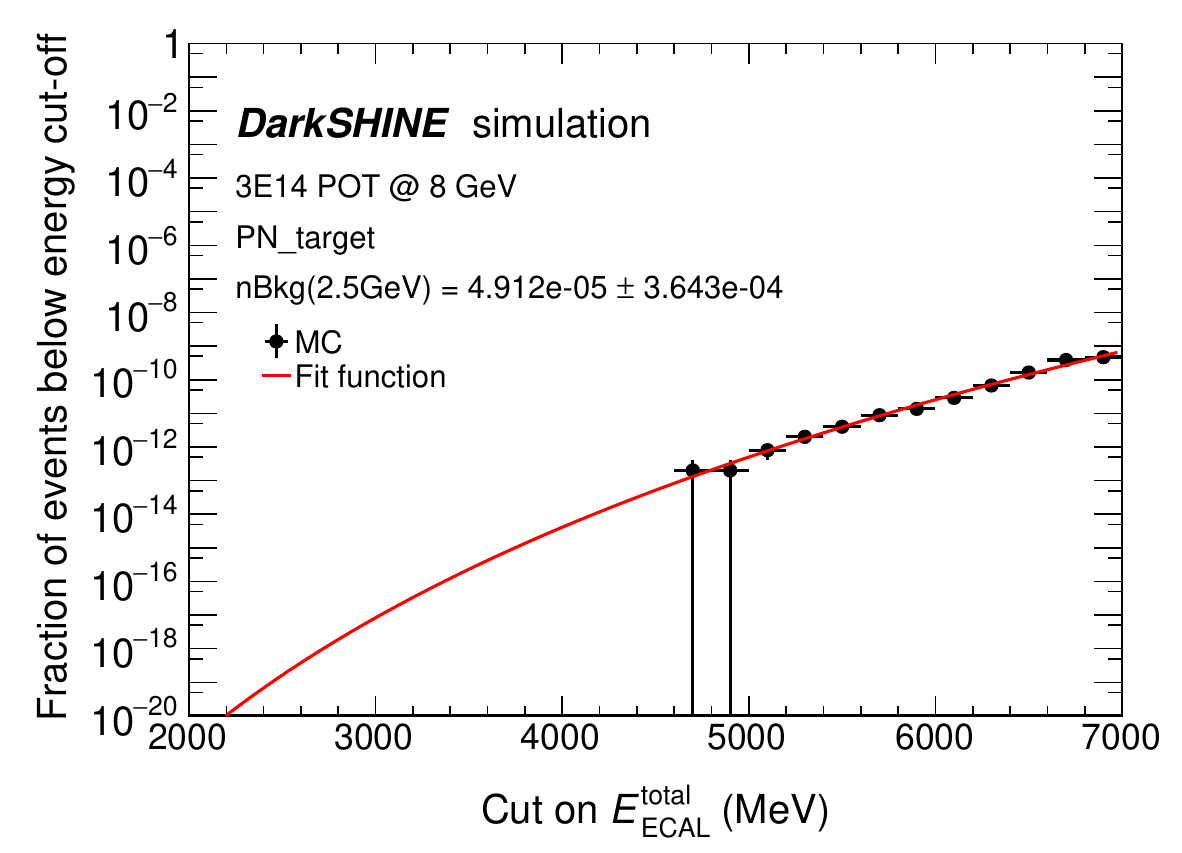}
\caption{Background extrapolation for signal region $e^++E_{\mathrm{miss}}$, the fraction of events below energy cutoff vs. ECAL energy.}
\label{fig:fit_rare_SR-Brem}
\end{figure}

For the signal region $e^{+}+E_{\mathrm{miss}}$ described in Chapter~\ref{sec:signal_region}, the extrapolation results of rare processes are shown in \Cref{fig:fit_rare_SR-Brem}, except for the GMM process, and the fractions in each rare process are scaled according to their branching ratio. Each rare process applied all cuts in signal region $e^{+}+E_{\mathrm{miss}}$ except the cut $E_{\text{ECAL}}^{\text{total}}$, then an exponential log function was used to fit the event ratio and the background yields were obtained at $E_{\text{ECAL}}^{\text{total}}$. As shown in the \Cref{tab:cutflow_Brem}, no GMM events survive the signal region selection criteria in current statistics. The ECAL total energy cut are not sensitive to the GMM process as shown in \Cref{fig:validation} because most of the energy is carried away by the muon pair and deposits into the HCAL. But this GMM process can be rejected by HCAL requirements ($E_{\text{HCAL}}^{total}$ and $E_{\text{HCAL}}^{MaxCell}$) and the remaining GMM events is less than the order of $10^{-6}$. Therefore, the contribution of the GMM process is neglected in this analysis and the background yields in signal region $e^++E_{\mathrm{miss}}$ is $0.1$ by summing up the extrapolated results of each rare process. 

\begin{figure}[htbp]
\centering
\includegraphics[width=.4\textwidth]{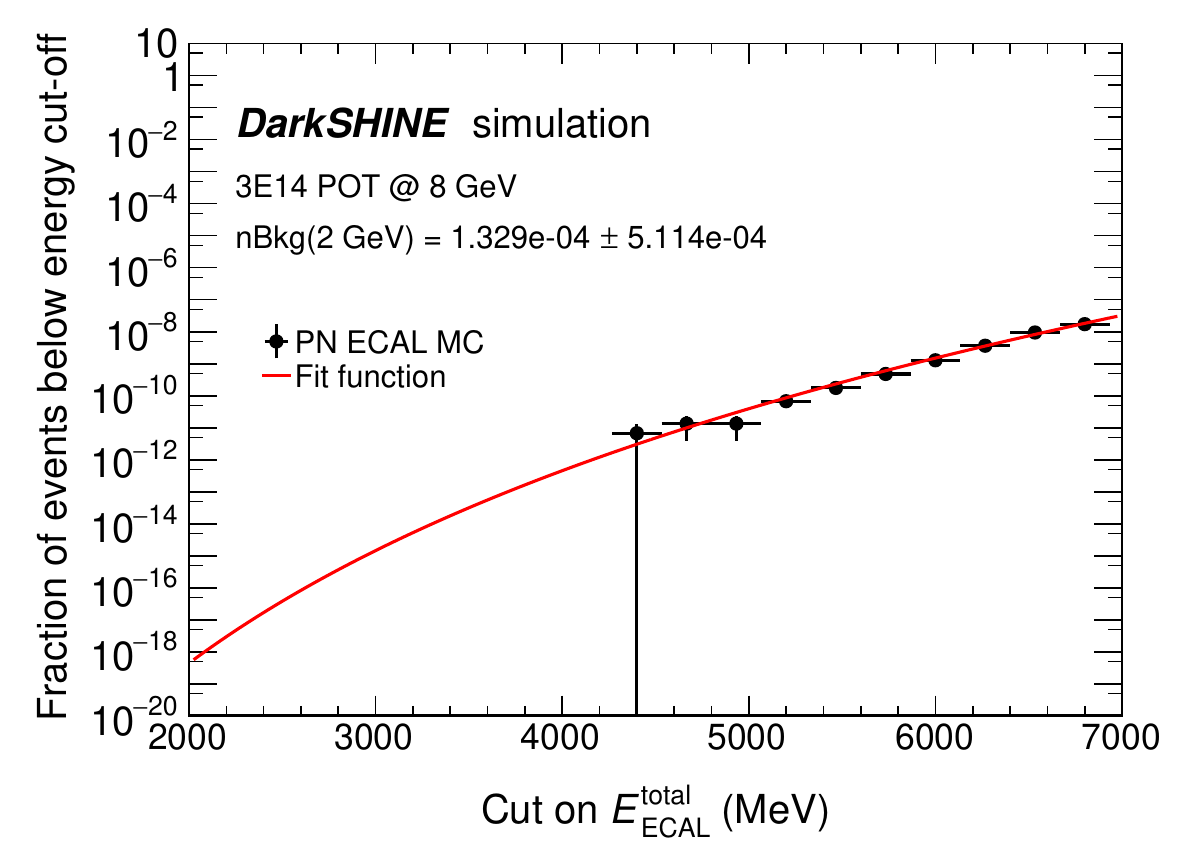}
\includegraphics[width=.4\textwidth]{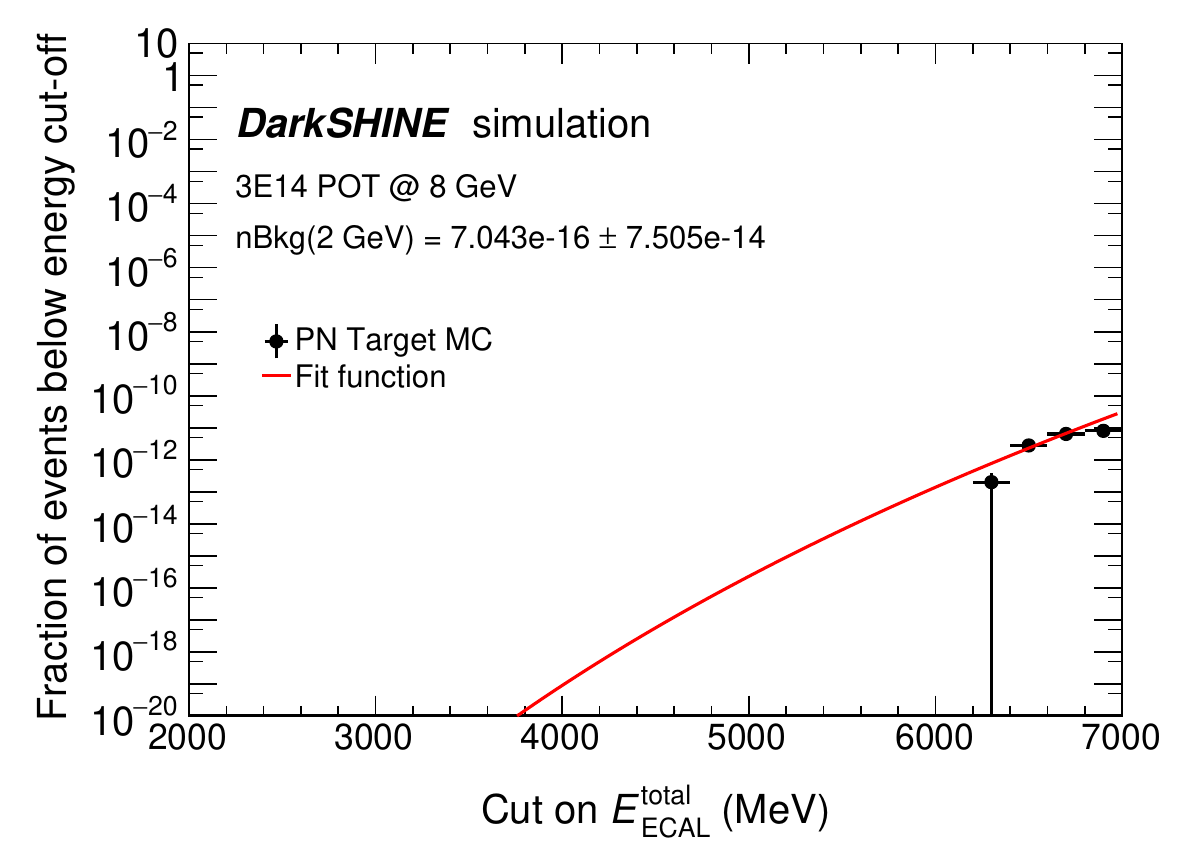}
\caption{Background extrapolation for signal region $\gamma+E_{\mathrm{miss}}$, the fraction of events below energy cutoff vs. ECAL energy.}
\label{fig:fit_rare_SR-NR}
\end{figure}

For signal region ($\gamma + E_{miss}$), the EN\_target process can be rejected by requiring the number of reconstructed track in recoil tracker is $0$. The background events from EN\_target process can be suppressed at the order of $10^{-6}$. As discussed above, the GMM process was negligible because of the effective HCAL requirements. Therefore, only the PN process remained in this signal region. The extrapolation results is shown in \Cref{fig:fit_rare_SR-NR}, where the fit function is also an exponential log function and the functional forms and parameters are adjusted accordingly. Finally, the total background yields is $0.016$ with the same extrapolation method as for signal region $e^++E_{\mathrm{miss}}$ at $E_{\text{ECAL}}^{\text{total}} = 2$ GeV.

\begin{figure}[htbp]
\centering
\includegraphics[width=.4\textwidth]{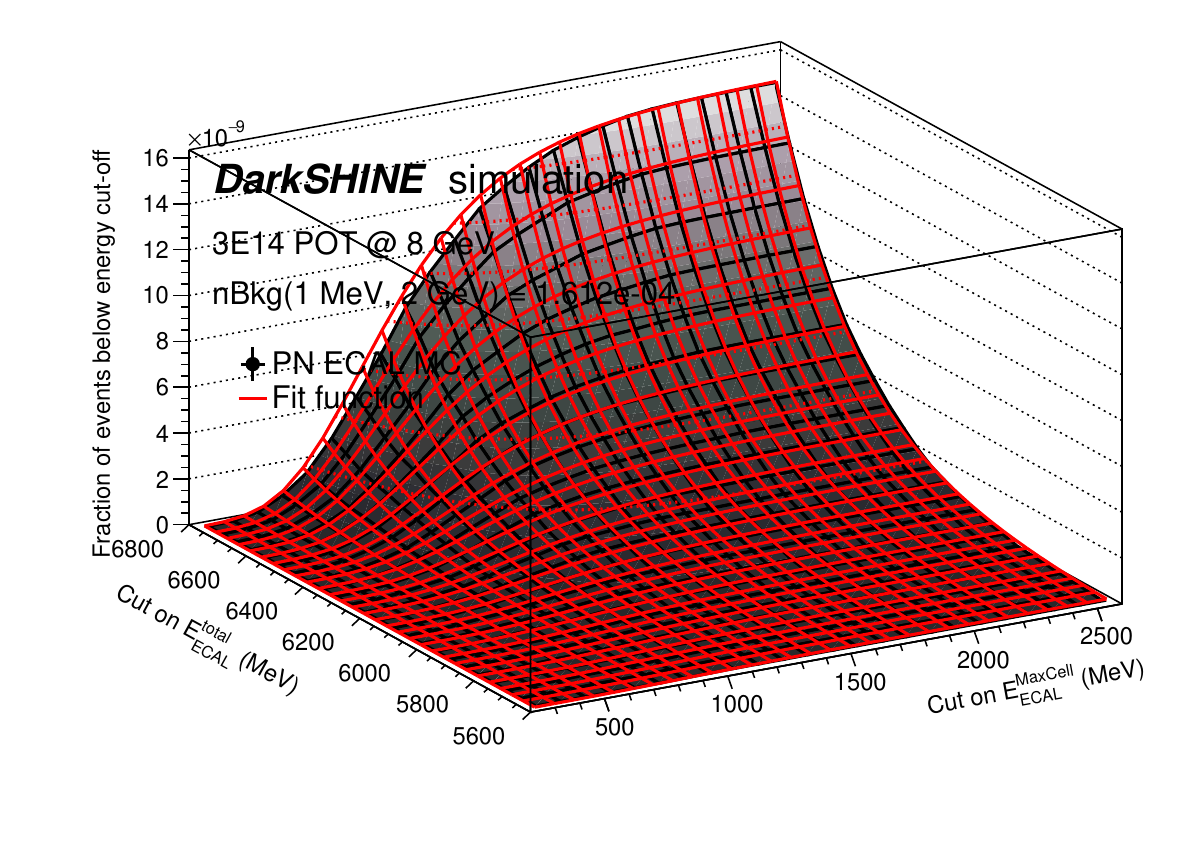}
\includegraphics[width=.4\textwidth]{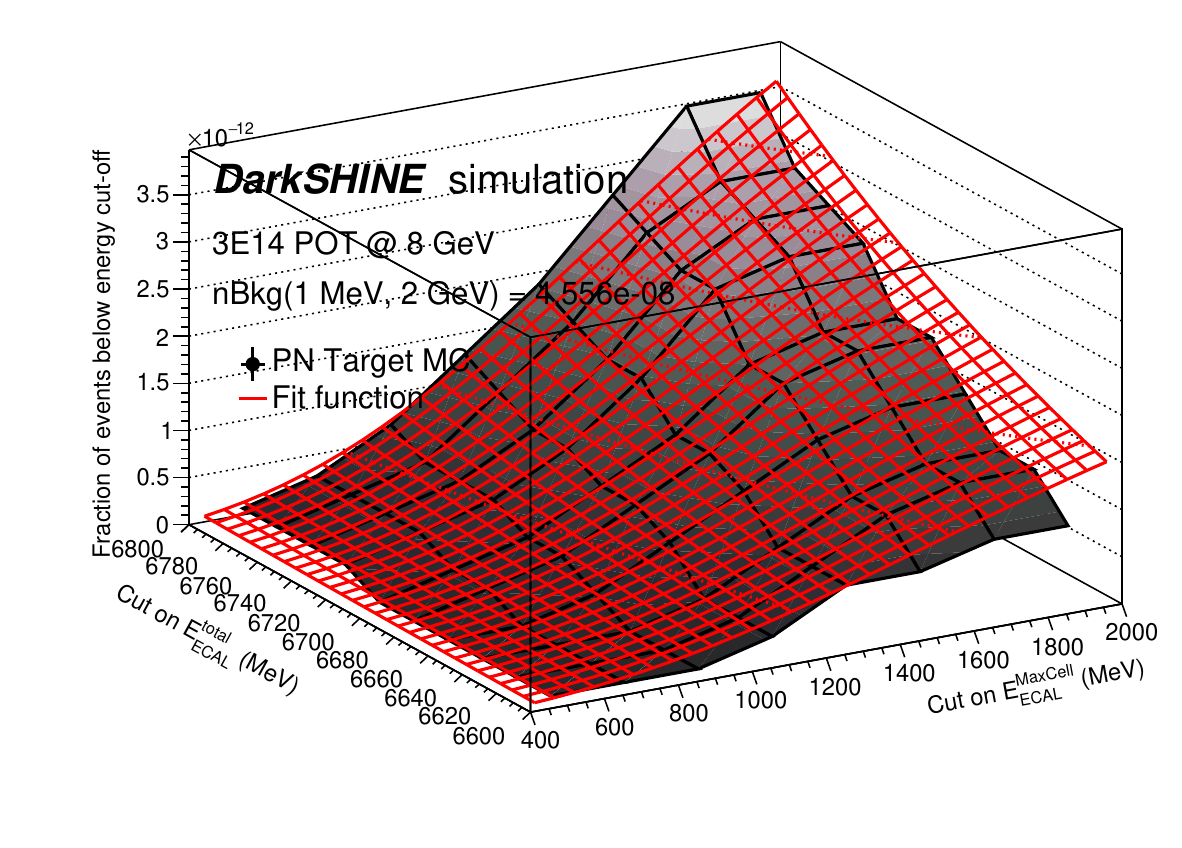}
\caption{Background extrapolation for signal region $E_{\mathrm{miss}}$, the fraction of events below energy cutoff vs. ECAL energy.}
\label{fig:fit_rare_SR-Res}
\end{figure}

For Signal region $E_{\mathrm{miss}}$, there are only the PN process since other rare processes were rejected or neglected as discussed above. \Cref{fig:fit_rare_SR-Res} shows the fit extrapolations for PN\_ECAL and PN\_target process. In this fit, all the cuts in signal region $E_{\mathrm{miss}}$ were applied except $E_{\text{ECAL}}^{\text{MaxCell}}$. The fitted function $e^{(a-be^{-cx})}$ shown in the plots as the red solid line, which can describe the shape of the event ratio. The total background yields in this signal region is $1.18$ by requiring $E_{\text{ECAL}}^{\text{MaxCell}}<1$ MeV.

\subsection{Validation from inclusive background simulation}
\label{sec:estimateBkg_inclusive}
To validate the background yields derived from rare process samples, we performed independent extrapolations using the inclusive background sample for each signal region. This cross-check ensures the consistency of our background modeling across different production mechanisms.

\begin{figure}[htbp]
\centering
\includegraphics[width=0.48\textwidth]{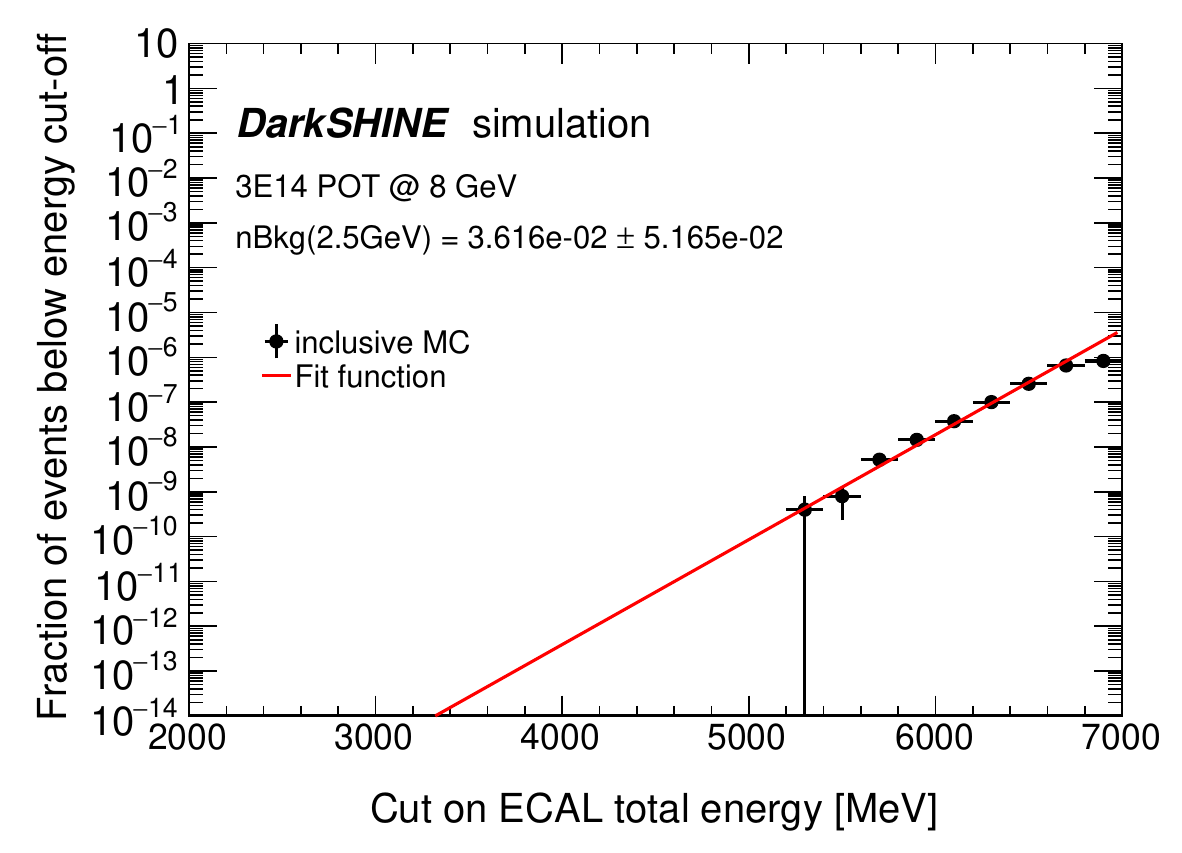}
\caption{Signal region $e^++E_{\mathrm{miss}}$ validation: Inclusive background extrapolation (blue circles) vs. rare process extrapolation (red triangles). Dashed lines show 68\% CL uncertainty bands.}
\label{fig:incl_fit_SRbrem}
\end{figure}

For signal region $e^++E_{\mathrm{miss}}$, we applied all selection criteria except the $E_\text{ECAL}^\text{total}$ requirement to the inclusive sample containing $2.5 \times 10^9$ POT events as shown in \Cref{fig:incl_fit_SRbrem} and the event survival fraction as a function of $E_\text{ECAL}^\text{total}$ cut. An exponential function $e^{ax^2+bx+c}$ was fit to these datasets. The inclusive extrapolation predicts $0.12 \pm 0.03$ background events, consistent with the rare process estimate of $0.10 \pm 0.02$.

\begin{figure}[htbp]
\centering
\includegraphics[width=0.48\textwidth]{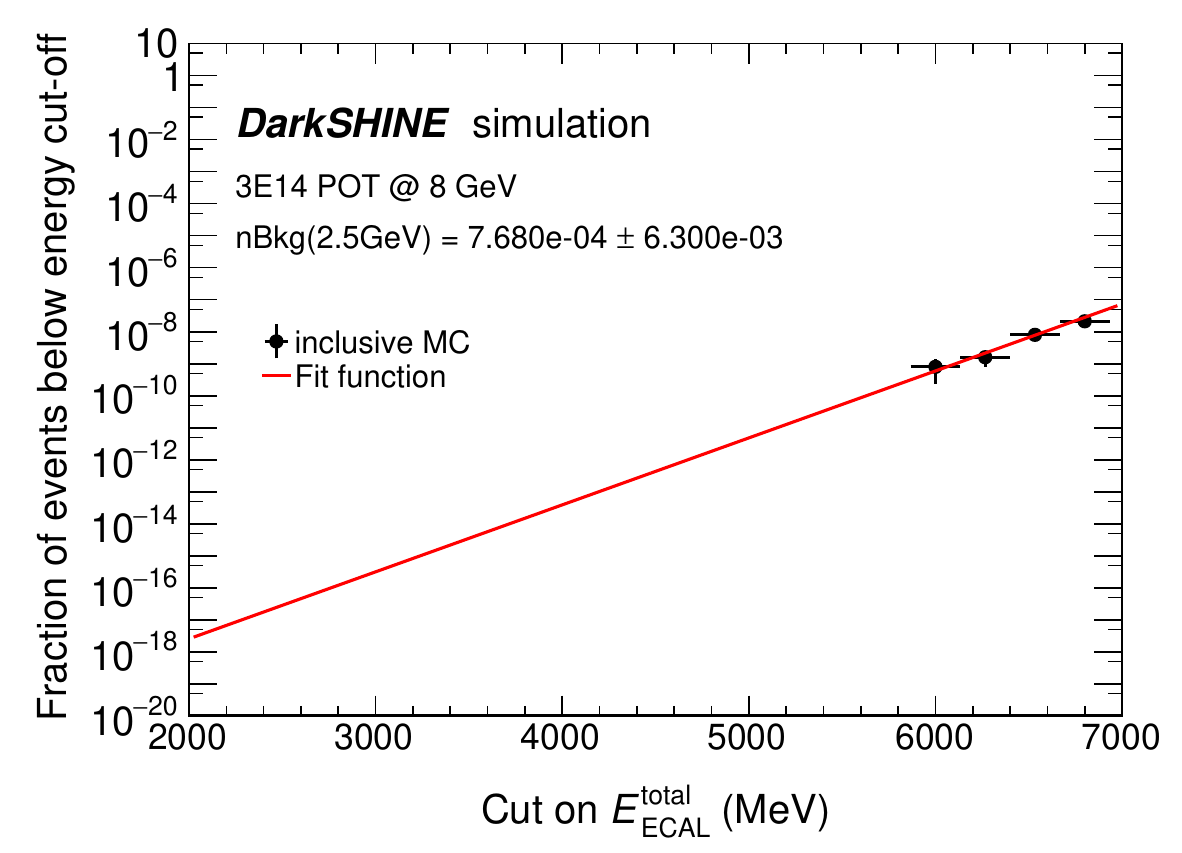}
\caption{Signal region $\gamma+E_{\mathrm{miss}}$ validation: Maximum ECAL cell energy distribution comparison between inclusive background (blue) and rare processes (red hatched).}
\label{fig:incl_fit_SRnr}
\end{figure}

For signal region $\gamma+E_{\mathrm{miss}}$, \Cref{fig:incl_fit_SRnr} demonstrates the fit result of extrapolation with inclusive background. In this fit, the all selection criteria in $\gamma+E_{\mathrm{miss}}$ except the $E_\text{ECAL}^\text{total}$ cut were applied and the fit function is $e^{-b(x - c)}$. The inclusive sample predicts $0.018 \pm 0.006$ events at $E_\text{ECAL}^\text{total}>2$ GeV, matching the rare process estimate of $0.016 \pm 0.005$.

\begin{figure}[htbp]
\centering
\includegraphics[width=0.48\textwidth]{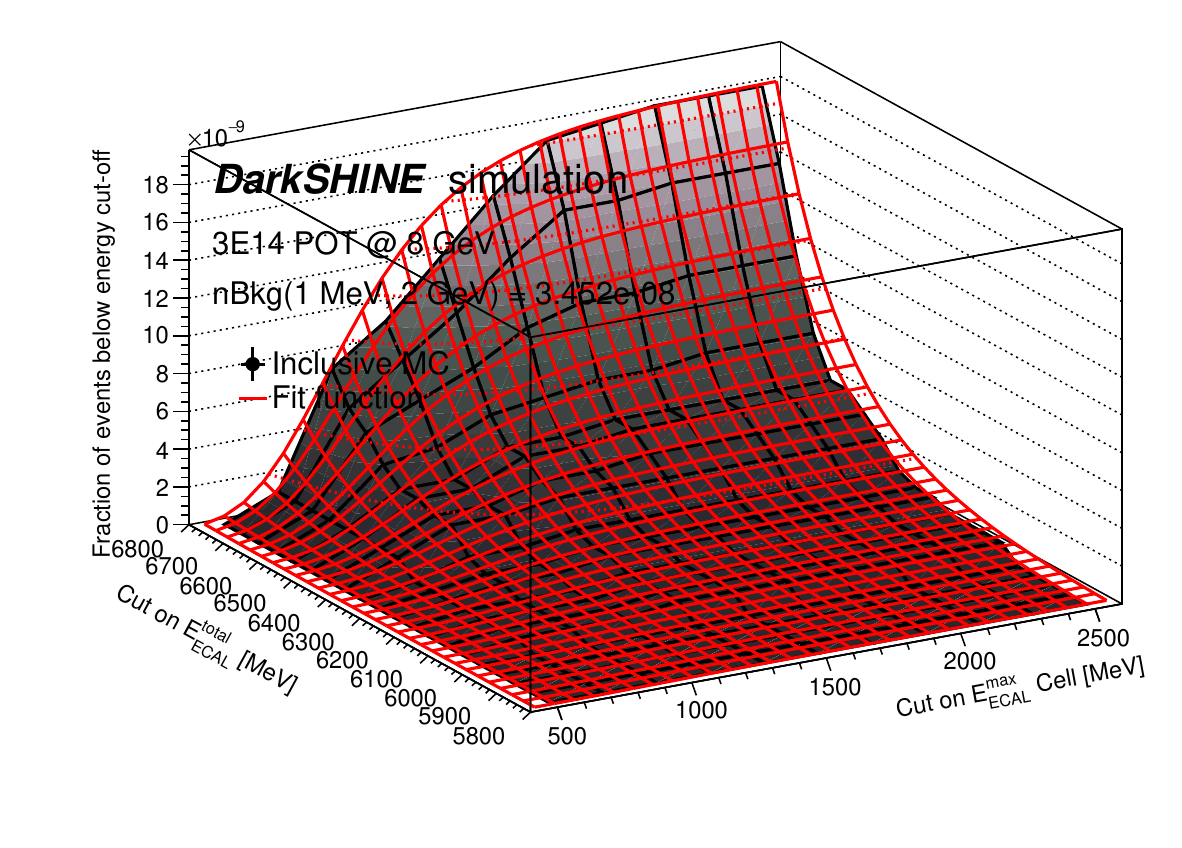}
\caption{Signal region $E_{\mathrm{miss}}$ validation: 2D distribution of maximum ECAL cell energy vs HCAL total energy in inclusive background. The red box indicates the signal region.}
\label{fig:incl_SRres_scatter}
\end{figure}

In the case of signal region $E_{\mathrm{miss}}$, \Cref{fig:incl_SRres_scatter} shows the extrapolation results with the inclusive sample and the function $e^{(a-be^{-cx})}$ was used to extrapolate into the signal region. The inclusive extrapolation yields $1.22 \pm 0.15$ events, in agreement with the rare process estimate of $1.18 \pm 0.12$.

The consistent agreement between inclusive and rare process extrapolations across all three SRs validates our background estimation methodology and the three background yields obtained from the rare background process in three corresponding signal region are used for the sensitivity study.

\section{Sensitivity study}

The relationship between the kinetic mixing parameter $\epsilon$ and the expected signal yield $N_{\text{sig}} = \sigma n L$ is calculated by CalcHEP, where $\sigma$ is the dark photon production cross-section, $L = 3\times10^{14}$ represents the number of positrons-on-target per run, and $n$ is the area density of target particles. For the bremsstrahlung production mechanism, $n$ is the atom density of tungsten target. For t-channel and s-channel production involving positron-electron scattering, $n$ is the electron number density of tungsten target.

The cross-section $\sigma$ spans orders of magnitude ($\sim 10^{2}\varepsilon^2$ to $\sim 10^{13}\varepsilon^2$ pb) across the 1–2000 MeV dark photon mass range. Asymptotic approximation is used for limit setting. A binned statistical analysis is performed using a profile likelihood ratio test statistic:
\begin{equation}
\lambda(\mu) = \frac{\mathcal{L}(\mu, \hat{\hat{\theta}})}{\mathcal{L}(\hat{\mu}, \hat{\theta})}
\end{equation}
where $\mu$ is the signal strength modifier and $\theta$ denotes nuisance parameters. The signal model incorporates three production channels (brem, $s$-ch, $t$-ch) with a shared normalization parameter $\mu$, while background shape uncertainties are modeled fully uncorrelated across bins. For each mass hypothesis:
\begin{equation}
\mu = \frac{\sigma}{\sigma_{\text{ref}}} \quad \text{with} \quad \sigma_{\text{ref}} \propto \varepsilon^2
\end{equation}
Expected limits at 90\% CL are derived using the $\mathrm{CL}_{\mathrm{s}}$ prescription under the background-only hypothesis ($\mu=0$). The nuisance parameters $\theta$ account for background normalization uncertainties derived from Poisson statistics ($\sqrt{N_{\text{bkg}}}$). A combined fit across all signal regions ($e^++E_{\mathrm{miss}}$, $\gamma+E_{\mathrm{miss}}$, $E_{\mathrm{miss}}$) is performed to obtain the overall search sensitivity.

\begin{figure}[htbp]
    \centering
    \includegraphics[width=0.8\textwidth]{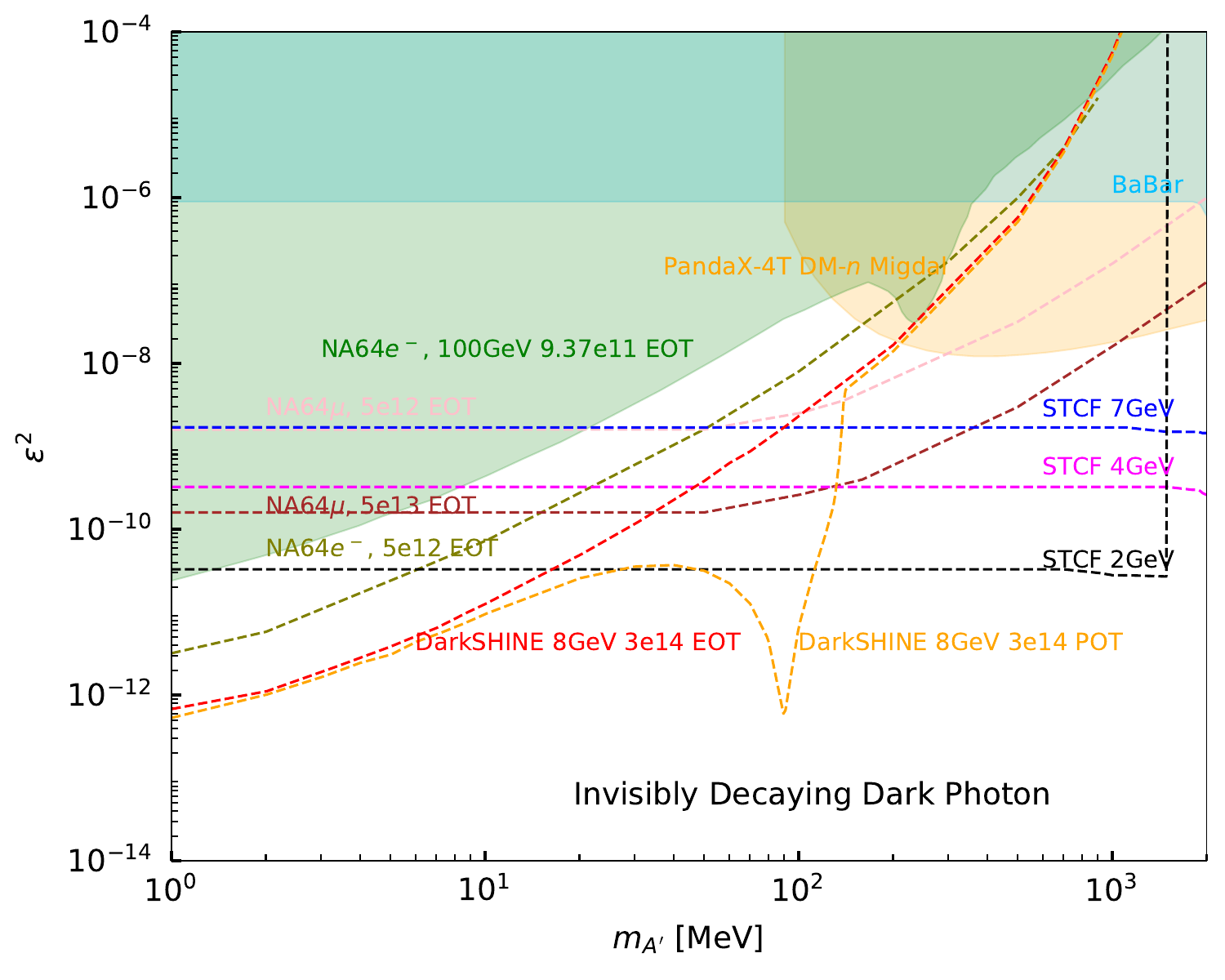}
    \caption{The new exclusion limits in ($\varepsilon^2$, $m_{A^{\prime}}$) plane for positron beam on the further DarkSHINE experiment, estimated with $3\times10^{14}$ POTs (yellow). The limits are shown in comparison for electron beam on the further DarkSHINE experiment with $3\times10^{14}$ EOTs (red), $9\times10^{14}$EOTs(orange), $1.5 \times 10^{15}$ EOTs (blue), and $10^{16}$ EOTs (green). The existing constraints on $\epsilon^2$ from the NA64~\cite{NA64:2023wbi}, PandX-4T~\cite{PandaX:2023xgl} and BaBar~\cite{BaBar:2017tiz} experiments are shown as a reference. In addition, the STCF sensitivity curves shown in the plot are computed assuming 30 $ab^{-1}$ at $\sqrt{s} = 7$ GeV, $\sqrt{s} = 4$ GeV, and $\sqrt{s} = 2$ GeV, respectively~\cite{Zhang:2019wnz}. The expected curves from the future NA64 experiments are also shown in the plot.}
    \label{fig:limitsCurve}
\end{figure}

The background number is based on the sum of extrapolated background yields in each signal region, as described in \Cref{sec:estimateBkg_rare}. The signal efficiencies were studied in Chapter~\ref{sec:signal_region}. We obtain the 90\% C.L. sensitivity illustrated by the yellow line in the \Cref{fig:limitsCurve}. The results suggest that could improve sensitivity by nearly two or more orders of magnitudes compared with the existing experiments, especially around $m_{A^{\prime}}=90$ MeV. 

\begin{figure}[htbp]
    \centering
    \includegraphics[width=0.8\textwidth]{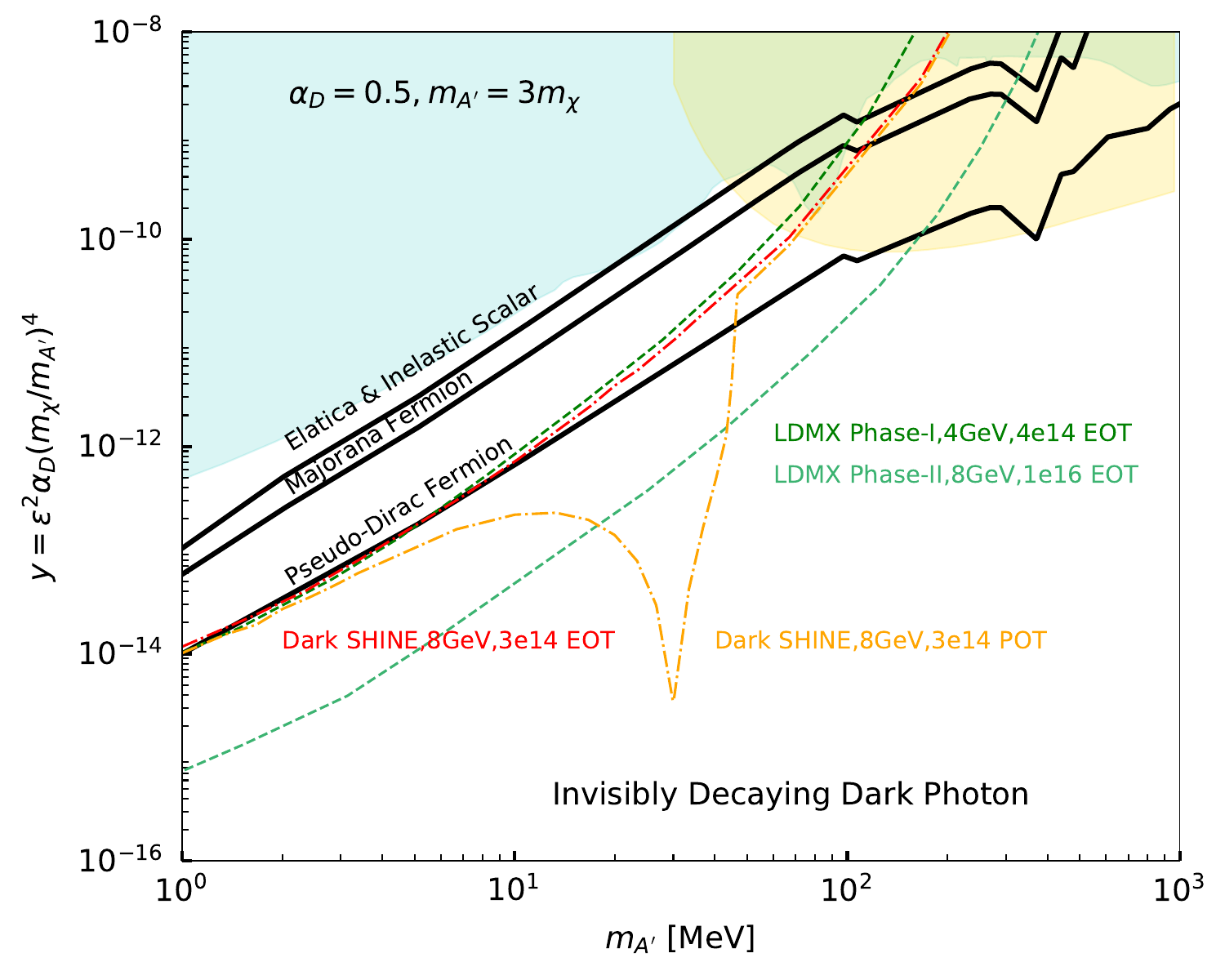}
    \caption{The new exclusion limits in ($y$, $m_{\chi}$) plane for positron beam on the future DarkSHINE experiment, obtained for $m_{A^{\prime}} = 3m_{\chi}$ and $\alpha_{D} = 0.5$, with $3\times10^{14}$ EOTs (yellow). For comparison, the limits for electron beam on the further DarkSHINE experiment with $3\times10^{14}$ EOTs (red), $9\times10^{14}$EOTs(blue), $1.5 \times 10^{15}$ EOTs (orange), and $10^{16}$ EOTs (pink) are shown in the plot. The turquoise region represents existing limits obtained in Ref.~\cite{NA64:2023wbi} from the results of the LSND~\cite{deNiverville:2011it,Batell:2009di},E137~\cite{Batell:2014mga}, BaBar~\cite{BaBar:2017tiz}, MiniBooNE~\cite{MiniBooNE:2017nqe}, COHERENT~\cite{COHERENT:2021pvd} and direct detection~\cite{Essig:2012yx} experiments. The latest result from PandX-4T experiment~\cite{Andreev:2021fzd} is also shown as the origin region in this plot. The projected limit from the LDMX experiment with $4 \times 10^{14} $EOTs with 4 GeV beam energy (Phase 1) and $10^{16}$ EOTs with 8 GeV beam energy (Phase 2) is shown in the plot as the green dashed curves~\cite{Akesson:2022vza}. The favored parameters for the observed relic DM density for the scalar, Majorana, and pseudo-Dirac of light thermal DM are shown as the solid curves~\cite{NA64:2017vtt}. } 
    \label{fig:dimensionlessLimitsCurve}
\end{figure}

To investigate thermal relic dark matter, \Cref{fig:dimensionlessLimitsCurve} presents the projected sensitivity of the dimensionless interaction strength $y=\epsilon^{2}\alpha_{D}(m_{\chi} / m_{A^{\prime}})^{4}$ as a function of DM mass $m_{\chi}$. These results is assumed that the mass of the dark photon $m_{A^{\prime}}$ is three times as large as the DM mass $m_{\chi}$ and that the coupling constant $\alpha_{D}$ between the dark photon $A^{\prime}$ and the DM $\chi$ is equal to 0.5. Three benchmark thermal targets (elastic and inelastic scalar, Majorana fermion, and pseudo-Dirac fermion) are shown as solid lines. The filled regions represent existing and projected experimental constraints~\cite{Andreev:2021fzd,BaBar:2017tiz,deNiverville:2011it,Batell:2009di,Batell:2014mga,MiniBooNE:2017nqe,PandaX:2023xgl,Essig:2012yx}. The results indicate that DarkSHINE’s sensitivity could probe thermal relic dark matter in the MeV range especially around $m_{A^{\prime}}=90$ MeV. These results demonstrate the significant potential of the DarkSHINE experiment to explore dark sector physics in positron beam mode of the SHINE facility. The observed sensitivity enhancements, particularly in the resonant annihilation regime, motivate the development of future positron beam at SHINE and moreover a variable beam energy setup will extend the search sensitivity competitiveness in a broader dark photon mass range. Such a program would leverage the distinctive features of $A^{\prime}$ production via positron annihilation to systematically map the light dark matter (LDM) parameter space, thereby enabling a comprehensive probe of coupling strengths and mass relationships within the dark sector.

\section{Summary}
In conclusion, we have performed the first prospective simulation analysis with the future DarkSHINE experiment setup in positron-on-fixed-target scheme to search for dark photon invisible decay signal. The resulting limits improved our last result by approximately two orders of magnitudes in electron-on-fixed-target scheme especially around $m_{A^{\prime}}=90$ MeV~\cite{DarkSHINE:2022mak}. This strongly motivates the promising physics potential of DarkSHINE to probe dark sectors in positron beam mode at SHINE facility and indicates a variable positron beam energy setup may even further help to more thoroughly search for $A^{\prime}$ production processes and more broadly scan the LDM parameter space for dark photon with the presence of its annihilation production channels in positron beam mode.


\acknowledgments
The work of X.L.Z., S.L., K.L., H.M.A., T.S., H.J.Y. and Y.F.Z. is supported by National Key R\&D Program of China (Grant No.: 2023YFA1606904 and 2023YFA1606900), National Natural Science Foundation of China (Grant No.: W2543005 and 12150006) and Shanghai Pilot Program for Basic Research—Shanghai Jiao Tong University (Grant No.: 21TQ1400209).



\bibliographystyle{JHEP}
\bibliography{biblio.bib}

\end{document}